\documentclass[12pt]{article}

\pdfoutput=1
\usepackage{amssymb}
\usepackage{amsmath}
\usepackage{latexsym}
\usepackage[usenames]{color}
\usepackage{cite}
\usepackage{refstyle}
\usepackage{diagbox}
\usepackage{setspace}
\usepackage[margin=2.0cm]{geometry}
{}

\usepackage{float}
\usepackage{hyperref}

\begin{document}

\begin{center}
$$$$
{\Large\textbf{\mathversion{bold}
 Complexity for superconformal primaries from BCH techniques 
}\par}
\vspace{1.0cm}

\textrm{ \textbf{Phumudzo Rabambi} $\,{}^{a\dagger}$ \ ,   \textbf{Hendrik J.R. van Zyl} $\,{}^{bc\ddag}$ }
\\ \vspace{1.5cm}

\begin{flushleft}
\emph{ ${}^a$ Mandelstam Institute for Theoretical Physics, School of Physics, NITheCS, and CoE-MaSS, University of the Witwatersrand, Johannesburg, WITS 2050, South Africa.}
\\ \vspace{3mm}

\emph{ ${}^b$The Laboratory for Quantum Gravity \& Strings, Department of Mathematics \& Applied Mathematics, University of Cape Town, Cape Town, South Africa}
\\ \vspace{3mm}

\emph{ ${}^c$The National Institute for Theoretical and Computational Sciences, Private Bag X1, Matieland, South Africa}
\\ \vspace{3mm}

${}^\dagger$\href{mailto:teflon.ac.za@gmail.com}{teflon.ac.za@gmail.com},$\,{}^*$ $\,{}^\ddag$\href{mailto:hjrvanzyl@gmail.com}{hjrvanzyl@gmail.com}\\

\end{flushleft}

\bibliographystyle{JHEP}

\par\vspace{1.5cm}

\textbf{Abstract}\vspace{2mm}
\end{center}
{
We extend existing results for the Nielsen complexity of scalar primaries and spinning primaries in four dimensions by including supersymmetry.  Specifically, we study the Nielsen complexity of circuits that transform a superconformal primary with definite scaling dimension, spin and R-charge by means of continuous unitary gates from the $\mathbf{\mathfrak{su}}(2,2|\mathcal{N})$ group.   Our analysis makes profitable use of Baker-Campbell-Hausdorff formulas including a special class of BCH formulas we conjecture and motivate.  With this approach we are able to determine the super-K\"{a}hler potential characterizing the circuit complexity geometry and obtain explicit expressions in the case of $\mathcal{N}=1$ and $\mathcal{N}=2$ supersymmetry. 

}

\setcounter{page}{1}
\newpage

\section{Introduction}

The key objective of quantum computational complexity is to quantify how hard it is to obtain a desired target state from a (typically simple) reference state by acting with a set of allowed (or accessible) unitary operations.  Much of the nomenclature present in these studies owe their origin to quantum computing where the unitary operators play the role of quantum gates and the reference and target state the input and output state of quantum circuits.  Quantum computational complexity has received significant recent attention within high energy physics due to its link with black hole geometry \cite{Susskind:2014moa,Susskind:2018pmk} through the AdS/CFT correspondence \cite{Maldacena:1997re}.  To be more precise, complexity is related to the growth of black hole interiors and the response of complexity to perturbations can be related to the response of the black hole interior to perturbations \cite{Brown:2015lvg, Brown:2015bva, Auzzi:2022bfd, Stanford:2014jda, Chapman:2018dem, Chapman:2018lsv}.  \\ \\
Due to this connection to black holes within the AdS/CFT correspondence it is clearly important to study complexity in the context of conformal field theories.  It has been shown, for free and weakly coupled quantum field theories, that complexity becomes the length of the shortest geodesic in the space of circuits \cite{Jefferson:2017sdb, Chapman:2017rqy, Khan:2018rzm, Hackl:2018ptj, Chapman:2018hou, Doroudiani:2019llj, Bhattacharyya:2018bbv, Jiang:2018nzg, Caceres:2019pgf, Guo:2020dsi, Meng:2021wmz, Moghimnejad:2021rqe, Yang:2017czx, Sinamuli:2019utz}, leading to a deep connection between geometry and complexity \cite{Brown:2019whu,Auzzi:2020idm}.  The presence of an enhanced symmetry in two dimensions has also lead to substantial results for the complexity of two-dimensional conformal field theories \cite{Caputa:2018kdj, Erdmenger:2020sup, Flory:2020eot, Flory:2020dja, Bueno:2019ajd}.  See also \cite{Bhattacharyya:2022ren,Ghodrati:2019bzz, Ghodrati:2017roz}, where complexity was studied in the context of warped conformal field theories.  \\ \\
It is also important to consider higher-dimensional cases to make contact with holographic theories in higher dimensions.  The paper \cite{Chagnet:2021uvi} has outlined a general approach for tackling this problem in $d$ dimensions.  The key idea, built on the formalism of Nielsen \cite{https://doi.org/10.48550/arxiv.quant-ph/0502070, Nielsen_2006, https://doi.org/10.48550/arxiv.quant-ph/0701004}, is to focus on a continuous notion of complexity, where a unitary gate may be constructed to be any group element of the $d$-dimensional conformal algebra.  The question of finding the computationally most efficient circuit connecting a reference and target state is then replaced by the question of finding the minimal geodesic connecting points on the manifold of quantum rays.  \\ \\
The restriction to the conformal symmetry group is a special choice, but one that comes with advantages.  The computational cost of synthesising a desired target state depends on the choice of cost function that penalises the use of certain unitary gates.  This choice is not unique.  However, when restricting to the conformal symmetry group and assuming all symmetry transformations to be equally easy to perform, the choice of cost function is fixed up to global choice of units \cite{Magan:2018nmu}.  This will also be the case for the symmetry groups that we consider and thus, by making this choice, we avoid the arbitrariness of the chioce of cost function.  Following this approach, \cite{Chagnet:2021uvi} determined the computational complexity for a scalar primary reference and target states in general dimensions.  In \cite{Koch:2021tvp} the computational complexity for spinning primaries was studied, with explicit results in three and four dimensions.  These works focused on the computation of the Fubini-Study metric, defined on the manifold of quantum rays, as cost function for computational complexity.   \\ \\
There is a further extension of the symmetry group that is relevant to the study of holography namely the inclusion of supersymmetry.  In this paper we will build on the work of \cite{Chagnet:2021uvi, Koch:2021tvp} to consider the computational complexity of states in representations of the superconformal algebra.   We will find it useful to compute the (super)-K\"{a}hler potential from the unnormalised overlaps of (super)-coherent states, following one of the approaches employed in \cite{Koch:2021tvp}.  The (super)-coherent state overlap is precisely the expectation value of a group element w.r.t. the reference state.  The group elements be manipulated by means of Baker-Campbell-Hausdorff formulae to obtain an explicit expression for the overlap.  To this end we conjecture a rather general BCH formula that reproduces several known results and that we have checked to high perturbative order.   For its application in this paper, manipulating exponentials of supercharge and conformal supercharge this formula is exact, owing to the presence of Grassmann variables.  This mathematical tool allows us to compute the (super)-coherent state overlap analytically for the $\mathcal{N}=1$ and $\mathcal{N}=2$ superconformal group in four dimensions.  With some modification it should also be applicable to superconformal theories in higher dimension and with additional supercharges. \\ \\ 
The paper is structured as follows.  We begin with a brief overview of circuit complexity, the FS metric and the circuits that we will be studying in section \ref{sectionCC}.  We then introduce the BCH formulae that are used to perform the computation including the conjectured formula and checks that we have performed in section \ref{sectionBCHformula}.  The BCH formulae are implemented in the derivation of the super-K\"{a}hler potentials for the $\mathcal{N}=1$ and $\mathcal{N}=2$ superconformal group in four dimensions in section \ref{sectionResults}.  We add some comments on how they may be applied in cases involving higher dimensions and additional supercharges.  We conclude with a discussion of the presented results and future directions in section \ref{discussion}.

\section{Circuit Complexity}
\label{sectionCC}

The aim of quantum computational complexity is to quantify how difficult it is to obtain a desired target state from a specified reference state, by applying some set of allowed unitary operations.  The Nielsen approach \cite{https://doi.org/10.48550/arxiv.quant-ph/0502070, Nielsen_2006, https://doi.org/10.48550/arxiv.quant-ph/0701004} considers the set of allowed unitary operations to be any element of some group $U \in G$.   The accessible target states are those that are related to the reference state by some unitary operation 
\begin{equation}
|\psi(\sigma) \rangle = U(\sigma) |\psi_0\rangle.  
\end{equation}
In the above the chosen reference state is $|\psi_0\rangle$ and $\sigma$ is the circuit parameter which is usually taken to run from $0$ to $1$.  \\ \\
The collection of accessible target states may thus be associated with the corresponding elements of the group $G$.  By defining a metric on the manifold of states one can quantify the computational complexity as distances on the manifold.  Specifically, the complexity is defined as the minimal geodesic length connecting the points corresponding to the reference and target state on the manifold.  As metric we follow the choice of \cite{Chagnet:2021uvi, Koch:2021tvp} namely the Fubini-Study metric
\begin{equation}
ds^2 = \langle \psi_0| d U^\dag dU |\psi_0 \rangle -  \langle \psi_0| d U^\dag U |\psi_0\rangle \langle \psi_0 | U^\dag dU |\psi_0 \rangle.    \label{FSMet}
\end{equation}
With this choice the distance between states differing by an overall phase is zero so that they should be identified with the same point on the manifold.  It is for this reason that the stability subgroup of $G$ should be highlighted.   This is the set of group elements $H \in G$ acting as
\begin{equation}
U_h |\psi_0 \rangle = e^{i \phi_h} |\psi_0\rangle     \label{stabilitySubGroup}
\end{equation}
on the chosen reference state.  The set of unitary transformations is thus more accurately described by elements of $G / H$.  Indeed, the target states we have in mind are in fact generalized coherent states \cite{Perelomov} and there is a one-to-one correspondence between points of the manifold of quantum  rays and group elements of $G / H$.  \\ \\
The coherent states (or equivalently the group elements of $G / H$) are labeled by some set of parameters $s \in (s_1, s_2, \cdots s_n$).  The FS-metric (\ref{FSMet}) may be computed directly \cite{Provost:1980nc} from the state overlap
\begin{eqnarray}
    ds^2 & = & g_{ij} ds^i ds^j   \nonumber \\
    g_{ij} &  = & \left. \partial_i \partial_j' \log | \langle s| s'\rangle |  \right|_{s' = s}   \label{FSState}
\end{eqnarray}
Note that any normalisation factor for the state does not contribute.  A special case arises when the (unnormalised) states are parameterised holomorphically by a set of complex coordinates $z = (z_1, z_2, \cdots z_m)$.   In this case the logarithm of the coherent state overlap is a K\"{a}hler potential \cite{Kriel:2015tga}
\begin{equation}
    g_{\bar{a} b}  = g_{\bar{b}a} =\frac{1}{2} \partial_a \partial_{\bar{b}} (z|z). \label{KahlerPot}
\end{equation} 
where $\partial_a = \frac{\partial}{\partial z_a}$ and $\partial_{\bar{a}} = \frac{\partial}{\partial \bar{z}_a}$ .  In \cite{Chagnet:2021uvi,Koch:2021tvp} a K\"{a}hler potential was obtained in the cases where a highest or lowest weight reference state was considered.  This is a consequence of the fact that, for a highest or lowest weight states,  either the up- or down-ladder operator is part of the stability subgroup (\ref{stabilitySubGroup}).  Since we are considering unitary actions, holomorphic and anti-holomorphic coordinates are paired with the up and down-ladder operators respectively.  The highest or lowest weight states may thus be parameterised holomorphically and the result (\ref{KahlerPot}) applies  
 \\ \\
The unitary operators we will be considering will be taken from the superconformal group in four dimensions, $\mathbf{\mathfrak{su}}(2,2|\mathcal{N})$. The generators, their algebra and hermiticity properties are unpacked in appendix \ref{AppendixA}.  The unitary operators we will be considering are of the form
\begin{equation}
U = e^{p^{ \dot{\alpha} \alpha} P_{\alpha \dot{\alpha} } } e^{q^\alpha_i Q^i_\alpha} e^{\bar{q}^{i \dot{\alpha}} \bar{Q}_{i \dot{\alpha}}  } e^{ k_{ \alpha \dot{\alpha}  } K^{ \dot{\alpha} \alpha } } e^{s_\alpha^i S_i^\alpha} e^{\bar{s}_{i \dot{\alpha}} \bar{S}^{i \dot{\alpha}}  }  e^{ d D} e^{ l^{\alpha}_\beta L_{\alpha}^{\ \beta}} e^{\bar{l}_{\dot{\alpha}}^{\dot{\beta}} \bar{L}^{\dot{\alpha}}_{\ \dot{\beta}} }   e^{ r_i^j R^i_j }  \label{Uparameters}
\end{equation}
where $q^\alpha_i, \bar{q}_{i \dot{\alpha}}, s_\alpha^i$ and $\bar{s}^{i \dot{\alpha}}$ are complex Grassmann variables and the other coefficients appearing inside the exponentials are complex variables.  We have made use of spinor indices that allow for a compact writing of the commutation relations, see appendix \ref{AppendixA}.  The Greek indices such  as $\alpha$ take values $\alpha = 1,2$, the dotted Greek indices $\dot{\alpha} = \dot{1}, \dot{2}$ while the Latin indices take values $i=1,2,\cdots \mathcal{N}$.  Repeated indices in super- and subscript are summed, a convention we will keep throughout.  Not all the variables appearing in the parameterisation (\ref{Uparameters}) are independent as some are determined by the unitary condition
\begin{equation}
    U^\dag U = I
\end{equation}
It will turn out that, for the highest weight representations we will consider, the unitarity constraints fix coefficients associated with the stability subgroup.     \\ \\
Our focus will be on the four-dimensional case where we may decompose the exponential of the rotation operators as
\begin{eqnarray}
e^{l^{\alpha}_\beta L_{\alpha}^{\ \beta}} & = & e^{ {l'}_{2}^1 L_{1}^{\ 2}} e^{ {l'}_0 L_1^{\ 1} } e^{ {l'}_{1}^2 L_{2}^{\ 1}} \nonumber \\
e^{ \bar{l}^{\dot{\beta}}_{\dot{\alpha}} \bar{L}^{\dot{\alpha}}_{\ \dot{\beta}}} & = & e^{ \bar{l}'^{\dot{2}}_{\dot{1}} \bar{L}^{\dot{1}}_{\ \dot{2} }}  e^{ \bar{l}'_0 \bar{L}_{\ \dot{1}}^{\dot{1}}} e^{ \bar{l}'^{\dot{1}}_{\dot{2}} \bar{L}^{\dot{2}}_{\ \dot{1}}} \nonumber
\end{eqnarray}
by means of $SU(2)$ Baker-Campbell-Hausdorff formulae. 
Our choice of reference state is made so as to maximise the number of generators that form part of the stability subgroup (\ref{stabilitySubGroup}) as this gives rise to the simplest possible form of the unitary circuits.  Our choice throughout will be a conformal primary of scaling dimension $\Delta$ and highest weight spins $h,\bar h$ i.e.
\begin{eqnarray}
D |\psi_0\rangle & = & \Delta |\psi_0\rangle \nonumber \\
K^{\alpha \dot{\alpha}} |\psi_0\rangle & = & 0 \nonumber \\
S_i^\alpha |\psi_0\rangle & = & 0 \nonumber \\
\bar{S}^{i \dot{\alpha}} |\psi_0\rangle & = & 0 \nonumber \\
L_{2}^{\ 2} |\psi_0\rangle & = & h |\psi_0\rangle \nonumber \\
L_{2}^{\ 1} |\psi_0\rangle & = & 0 \nonumber \\
\bar{L}_{\ \dot{1}}^{\dot{1} }|\psi_0\rangle & = & \bar{h} |\psi_0\rangle  \nonumber \\
\bar{L}^{\dot{2}}_{\ \dot{1}} |\psi_0\rangle & = & 0  \label{refState}
\end{eqnarray}
In addition to this we need to specify how the state transforms under the $R$-charge generators which we will do on a case-by-case basis.  Schematically, our reference state is thus 
\begin{equation}
   |\psi_0\rangle = |\Delta; h,h; \bar{h},\bar{h}; \left\{ R\right\} \rangle   \label{refStateExplicit}
\end{equation}
Up to an overall normalisation constant the target state becomes
\begin{equation}
|\psi_f(\sigma) \rangle  = N e^{ p^{\alpha \dot{\alpha}} P_{\dot{\alpha} \alpha} } e^{q^\alpha_i Q^i_\alpha} e^{\bar{q}^{ i\dot{\alpha}} \bar{Q}_{i \dot{\alpha}}  } e^{ l^{1}_2 L_{1}^{\ 2}} e^{ \bar{l}_{\dot{1}}^{\dot{2}} \bar{L}^{\dot{1}}_{\ \dot{2}} }   e^{ r_i^j R^i_j }  |\psi_0\rangle    \label{superCoherentStates}
\end{equation}
We still need to specify how the reference state transforms under the action of the R-charge generators. For $\mathcal{N}=1, 2$ we will still make a choice that simplifies the circuit the most.  Note that, if there is no supersymmetry present, the final state takes the form as studied in \cite{Koch:2021tvp}.  The states (\ref{superCoherentStates}) are generalised super-coherent states   \cite{Pelizzola:1992ab} of $\mathbf{\mathfrak{su}}(2, 2 | \mathcal{N})$.  \\ \\
We require a generalisation of the Fubini-Study metric to include the complex Grassmann-valued variables alongside the complex real variables.  The resulting expression is a natural extension of (\ref{FSState}) to a superspace metric \cite{Arnowitt:1975bd}.  As before the complexity is given by computing the minimal geodesic.  Our computational steps are as follows:  We start by computing the super-coherent state overlap
\begin{eqnarray}
& &  \langle \psi'_f(\sigma) |\psi_f(\sigma) \rangle \nonumber \\
&=& \langle \psi_{0} | e^{\bar{r}^j_i R^i_j } e^{ \bar{l}_{\dot{2}}^{\dot{1}} \bar{L}^{\dot{2}}_{\ \dot{1}} }  e^{ l^{2}_1 L_{2}^{\ 1}} e^{ \bar{s}_{i \dot{\alpha}}  \bar{S}^{i \dot{\alpha}}  } e^{ s^i_\alpha S^\alpha_i }   e^{ k_{ \dot{\alpha} \alpha }   K^{\alpha \dot{\alpha}} }   e^{ p^{\alpha \dot{\alpha}} P_{\dot{\alpha} \alpha} } e^{q^\alpha_i Q^i_\alpha} e^{\bar{q}^{i \dot{\alpha}} \bar{Q}_{i \dot{\alpha}}  } e^{ l^{2}_1 L_{1}^{\ 2}} e^{ \bar{l}_{\dot{1}}^{\dot{2}} \bar{L}^{\dot{1}}_{\ \dot{2}} }  e^{r^j_i R^i_j } |\psi_{0}\rangle \nonumber \\
& & \label{Ouroverlap}
\end{eqnarray}
To obtain the above we have made use of the conjugation relations (\ref{hermConj1}, \ref{hermConj2}).  Note that the coeffcients above are related as
\begin{equation}
    k_{\dot{\alpha} \alpha } = (p^{\alpha \dot{\alpha} })^* \ \ ; \ \ (s_\alpha^i) = (q_i^\alpha)^* \ \ ; \ \ \bar{s}_{i \dot{\alpha}} = (\bar{q}^{i \dot{\alpha}})^* \ \ ; \ \ l^2_1 = (l^1_2)^* \ \ ; \ \ \bar{l}_{\dot{2}}^{\dot{1}} = (\bar{l}_{\dot{1}}^{\dot{2}})^*  \ \ ; \ \ \bar{r}^j_i = (r^i_j)^*  \nonumber
\end{equation}
Since we will be making use of the highest weight representation reference state, the resulting super-coherent states will be parameterised holomorphically.  The logarithm of the super-coherent state overlap is, in fact, a super K\"{a}hler potential \cite{ElGradechi:1994wq} and the metric is given by taking appropriate derivatives
\begin{eqnarray}
ds^2 & = & \left( \sum_{j} d z_j \partial_{z_j}  +  \sum_{\beta}   d\theta_\beta   \partial_{\theta_\beta}  \right) \left( \sum_{i} d z_i^* \partial_{z_i^*}  +  \sum_{\alpha}   d\theta_\alpha^*   \partial_{\theta_\alpha^*}  \right)   \log \left( \langle \psi_f(\sigma) |\psi_f(\sigma) \rangle  \right)  \nonumber \\
& & 
\end{eqnarray}
w.r.t. the complex real and complex Grassmann variables.  \\ \\
For the purpose of this paper our focus will be on computing the super-K\"{a}hler potential.  Its form fully determines the resulting geometry which in turn determines the geodesic lengths that capture the computational complexity.   We leave a detailed study of these geometries to future work.  

\section{BCH formulae}
\label{sectionBCHformula}

The super-coherent state overlap may be computed in a variety of ways, but we have found the use of Baker-Campbell-Hausdorff formulas to "swap" exponentials of generators to be a powerful and efficient way to perform the computation.  This approach is also used in \cite{Koch:2021tvp} to compute the complexity for spinning primaries.  The expressions we have in mind start as the product of two exponentials
\begin{equation}
    e^{A} e^B     \nonumber
\end{equation}
where $A$ annihilates reference state ket and $B$ the reference state bra.  The "swap" then involves writing the above in the form 
\begin{equation}
    e^{A} e^B = e^{B'} e^{C} e^{A'} 
\end{equation}
where $A'$ ($B'$) annihilates the ket (bra).  The "swap" is thus an exchange of operators annihilating the ket (bra) from the left (right) to the right (left) of the product of exponentials. \\ \\
With this aim in mind, we now present the following special class of BCH formulas that, as far as we are aware, is a new mathematical result.  Our starting point is the following definitions of commutators
\begin{eqnarray}
C_{n+1} & = & \left[ \left[ A, C_{n}\right], B\right] \\
C_{1} & = & \left[ A, B \right] \nonumber \\
A_{n+1} & = & \left[ A_{1}, C_{n}\right]   \nonumber \\
B_{n+1} & = & \left[ C_n, B_{1} \right] \nonumber
\end{eqnarray}
where $A_{1} = A$ and $B_{1} = B$.  It immediately follows that 
\begin{equation}
C_{n+1} = \left[ A_n, B \right]   \label{CfromAnB}
\end{equation}
The \textbf{key assumptions} for this special class of BCH formulas are
\begin{eqnarray}
\left[ A_{i}, A_{j} \right] & = & 0   \nonumber \\
\left[ B_{i}, B_{j} \right] & = & 0   \label{Conditions} \\
\left[ C_{i}, C_{j} \right] & = & 0   \nonumber 
\end{eqnarray}
for all $i, j$.  Using these conditions, the definitions and the Jacobi identity we derive
\begin{eqnarray}
\left[ A_{i}, C_{j} \right] & = & A_{i+j} \nonumber \\
\left[ C_{j}, B_{i} \right] & = & B_{i+j}  \\
\left[ A_{i}, B_{j} \right] & = & C_{i+j-1} \nonumber
\end{eqnarray}
This represents a dramatic simplification of the commutator structure that arises from swapping powers of $A$ and $B$.  In this special case we find that the following identity holds
\begin{eqnarray}
e^{A} e^{B} & = & \prod_{j=1}^\infty e^{(\frac{1}{2})^{j-1} B_{j}} \ \prod_{j=1}^\infty e^{(\frac{1}{2})^{j-1} \frac{1}{j} C_{j}} \ \prod_{j=1}^\infty e^{(\frac{1}{2})^{j-1} A_{j}}   \nonumber \\
& = &  e^{ \sum_{j=1}^\infty (\frac{1}{2})^{j-1} B_{j}} \ e^{ \sum_{j=1}^{\infty} (\frac{1}{2})^{j-1} \frac{1}{j} C_{j}} \ e^{\sum_{j=1}^{\infty}(\frac{1}{2})^{j-1} A_{j}}     \label{newBCH}
\end{eqnarray}
One can verify the above term by term and this has been checked up to fifteenth order using a Mathematica code.  To perform this check, note that the operators $B_j$ contribute $j$ factors of $B$ and $j-1$ factors of $A$ and vice versa for $A_j$.  The factors $C_j$ contribute $j$ factors of $A$ as well as $j$ factors of $B$.  \\ \\
  A simple but instructive example of how to use this general formula to obtain specific BCH formulas is given in Appendix (\ref{RecursionAppendix}).  Once obtained, formulas may be verified using matrix representations or expanded order by order.   In the upcoming computations we will make use of three BCH formulas that may be derived from (\ref{newBCH}). The first two are the swap rules for supercharges and conformal supercharges are
\begin{eqnarray}
e^{s^j_\beta S_j^\beta} e^{q_i^\alpha Q^i_\alpha} & = & e^{\left( C_q(s, q) \right)_i^\alpha Q^i_\alpha} e^{C_d(s, q) D} e^{(C_l(s, q))_\beta^\alpha L_\alpha^{\ \beta}} e^{(C_r(s , q))_i^j R^i_j } e^{(C_s(s, q))^j_\beta S_j^\beta}     \label{SQSwap}   \\
e^{\bar{s}_{j \dot{\beta}} \bar{S}^{j \dot{\beta}} } e^{\bar{q}^{i \dot{\alpha}} \bar{Q}_{i\dot{\alpha}}} & = & e^{\left( \bar{C}_q(\bar{s}, \bar{q}) \right)^{i \dot{\alpha}} \bar{Q}_{i \dot{\alpha}}} e^{\bar{C}_d(\bar{s}, \bar{q}) D}  e^{(\bar{C}_{\bar{l}}(\bar{s}, \bar{q}))_{\dot{\alpha}}^{\dot{\beta}} \bar{L}^{\dot{\alpha}}_{\ \dot{\beta}}}e^{(\bar{C}_r(\bar{s} , \bar{q}))_i^j R^i_j } e^{(\bar{C}_s(\bar{s}, \bar{q}))_{j \dot{\beta}} \bar{S}^{j \dot{\beta}}} \label{SbQbSwap}
\end{eqnarray}
where 
\begin{eqnarray}
(C_q(s, q))_i^{\alpha} & = &  2\frac{2 - K_{+} - K_{-} }{(2 - K_{-})(2 - K_{+})}q_{i}^{\alpha} + \frac{4}{(2 - K_{-})(2 - K_{+})} q_i^\beta s^j_\beta q_j^\alpha   \nonumber \\
(C_s(s, q))_\beta^{j} & = &  2\frac{2 - K_{+} - K_{-} }{(2 - K_{-})(2 - K_{+})}s_{\beta}^{j} + \frac{4}{(2 - K_{-})(2 - K_{+})} s_\beta^k q^\gamma_k s_\gamma^j   \nonumber \\
C_d(s,q) & = & -\frac{1}{2}\log\left( \left(1 - \frac{K_{-}}{2} \right)\left(1 - \frac{K_{+}}{2} \right)   \right)   \nonumber \\
(C_{l}(s,q))^\alpha_\beta & = & 2\left( \frac{K_{+} \log\left( 1 - \frac{K_{-}}{2}\right)}{K_{-} (K_{-} - K_{+}) } -  \frac{K_{-}\log\left( 1 - \frac{K_{+}}{2}\right)}{K_{+} (K_{-} - K_{+}) } \right) s^{j}_\beta q^\alpha_j  \nonumber \\
& & - 4\left( \frac{ \log\left( 1 - \frac{K_{-}}{2}\right)}{K_{-} (K_{-} - K_{+}) } -  \frac{\log\left( 1 - \frac{K_{+}}{2}\right)}{K_{+} (K_{-} - K_{+}) } \right)s^{l}_\beta q_l^\gamma s_\gamma^k q_k^\alpha     \nonumber \\
(C_{r}(s,q))^j_i & = & -2\left( \frac{K_{+} \log\left( 1 - \frac{K_{-}}{2}\right)}{K_{-} (K_{-} - K_{+}) } -  \frac{K_{-}\log\left( 1 - \frac{K_{+}}{2}\right)}{K_{+} (K_{-} - K_{+}) } \right) s^{j}_\beta q^\beta_i  \nonumber \\
& & - 4\left( \frac{ \log\left( 1 - \frac{K_{-}}{2}\right)}{K_{-} (K_{-} - K_{+}) } -  \frac{\log\left( 1 - \frac{K_{+}}{2}\right)}{K_{+} (K_{-} - K_{+}) } \right)s^{j}_\gamma q_k^\gamma s_\beta^k  q_i^\beta \nonumber \\
K_{\pm }(s,q) &=& s_{\beta}^j q_{j}^\beta  \pm \sqrt{2 s_\beta^j q_j^\gamma s_\gamma^k q_k^\beta - (s_{\beta}^j q_{j}^\beta)^2 } \nonumber
\end{eqnarray}
and
\begin{eqnarray}
(\bar{C}_q(\bar{s}, \bar{q}))^{i \dot{\alpha}} & = &  2\frac{2 - \bar{K}_{+} - \bar{K}_{-} }{(2 - \bar{K}_{-})(2 - \bar{K}_{+})} \bar{q}^{i \dot{\alpha}} + \frac{4}{(2 - \bar{K}_{-})(2 - \bar{K}_{+})} \bar{q}^{i \dot{\beta}} \bar{s}_{j \dot{\beta}} \bar{q}^{j \dot{\alpha}}   \nonumber \\
(\bar{C}_s(\bar{s}, \bar{q} ))_{j \dot{\beta}} & = &  2\frac{2 - \bar{K}_{+} - \bar{K}_{-} }{(2 - \bar{K}_{-})(2 - \bar{K}_{+})}s_{j \dot{\beta}} + \frac{4}{(2 - \bar{K}_{-})(2 - \bar{K}_{+})} \bar{s}_{k \dot{\beta}} \bar{q}^{k \dot{\dot{\gamma}}} \bar{s}_{j \dot{\gamma}}   \nonumber \\
\bar{C}_d(\bar{s},\bar{q}) & = & -\frac{1}{2}\log\left( \left(1 - \frac{\bar{K}_{-}}{2} \right)\left(1 - \frac{\bar{K}_{+}}{2} \right)   \right)   \nonumber \\
(\bar{C}_{\bar{l}}(\bar{s},\bar{q}))^{\dot{\beta}}_{\dot{\alpha}} & = & 2\left( \frac{\bar{K}_{+} \log\left( 1 - \frac{\bar{K}_{-}}{2}\right)}{\bar{K}_{-} (\bar{K}_{-} - \bar{K}_{+}) } -  \frac{\bar{K}_{-}\log\left( 1 - \frac{\bar{K}_{+}}{2}\right)}{\bar{K}_{+} (\bar{K}_{-} - \bar{K}_{+}) } \right) \bar{s}_{j \dot{\alpha}} \bar{q}^{j \dot{\beta}}  \nonumber \\
& & - 4\left( \frac{ \log\left( 1 - \frac{\bar{K}_{-}}{2}\right)}{\bar{K}_{-} (\bar{K}_{-} - \bar{K}_{+}) } -  \frac{\log\left( 1 - \frac{\bar{K}_{+}}{2}\right)}{\bar{K}_{+} (\bar{K}_{-} - \bar{K}_{+}) } \right) \bar{s}_{l\dot{\alpha}} \bar{q}^{l \dot{\gamma}} \bar{s}_{k \dot{\gamma}} \bar{q}^{k \dot{\beta}}     \nonumber \\
(\bar{C}_{r}(\bar{s},\bar{q}))^j_i & = & 2\left( \frac{\bar{K}_{+} \log\left( 1 - \frac{\bar{K}_{-}}{2}\right)}{\bar{K}_{-} (\bar{K}_{-} - \bar{K}_{+}) } -  \frac{\bar{K}_{-}\log\left( 1 - \frac{\bar{K}_{+}}{2}\right)}{\bar{K}_{+} (\bar{K}_{-} - \bar{K}_{+}) } \right) \bar{s}_{i \dot{\beta}} \bar{q}^{j \dot{\beta}}  \nonumber \\
& & + 4\left( \frac{ \log\left( 1 - \frac{\bar{K}_{-}}{2}\right)}{\bar{K}_{-} (\bar{K}_{-} - \bar{K}_{+}) } -  \frac{\log\left( 1 - \frac{\bar{K}_{+}}{2}\right)}{\bar{K}_{+} (\bar{K}_{-} - \bar{K}_{+}) } \right) \bar{s}_{i \dot{\gamma}} \bar{q}^{ k \dot{\gamma}} \bar{s}_{k \dot{\beta} }  \bar{q}^{j \dot{\beta}} \nonumber \\
\bar{K}_{\pm }(\bar{s},\bar{q}) &=& \bar{s}_{j \dot{\beta}} \bar{q}^{j \dot{\beta} }  \pm \sqrt{2 \bar{s}_{j \dot{\beta}} \bar{q}^{j \dot{\gamma}} \bar{s}_{k \dot{\gamma}} \bar{q}^{k \dot{\beta}} - (\bar{s}_{j \dot{\beta}} \bar{q}^{j \dot{\beta}})^2 } \nonumber
\end{eqnarray}
At first glance the functions of Grassmann variables may seem problematic due to their anti-commuting nature.  However, note that, where they appear in functions above, the Grassmann-valued variables always appear in pairs as $s_{\alpha}^i q_j^\beta$ and $\bar{s}_{i \dot{\alpha}} \bar{q}^{j \dot{\beta}}$.  When expanding using these paired-up variables the series expansion can be done without ordering ambiguities appearing.  A useful feature of the above is that, due to the Grassmann-valued variables, the series expansions of (\ref{SQSwap}, \ref{SbQbSwap}) terminate at some finite order, allowing us to verify the swap rules explicitly.   \\ \\
In deriving the above rules we have worked in four dimensions so that $\alpha=1,2$ and $\dot{\alpha} =\dot{1},\dot{2}$ but we have not assumed any particular value of $\mathcal{N}$.  Indeed, the swap rules (\ref{SQSwap}), (\ref{SbQbSwap}) are valid provided that the number of Greek indices are $2$ \textbf{or} the number of Latin indices are $2$.  The intuitive way to understand this is that, since all indices appearing in the expressions are contracted, one may introduce intermediate variables where the Latin or Greek indices are contracted.  These intermediate variables may be packaged into matrices.  The matrix indices also need fully contracted so that the relevant quantities are thus traces of products of these matrices.   The trace of a $2\times 2$ matrix raised to an arbitrary power may be written as a function of the trace of the matrix and the trace of the matrix squared.  Thus only two fully contracted variables, namely the trace of the matrix and the trace of the matrix squared, appear in the expressions - these are closely related to the functions $K_{+}, K_{-}$ that appear above.  \\ \\ 
In four dimensions the exponentials in (\ref{SQSwap}), (\ref{SbQbSwap}) involving the rotation operators can be decomposed further using $SU(2)$ BCH formulas.  These give
\begin{eqnarray}
& & exp\left\{ (C_l)^{\alpha}_\beta L_{\alpha}^{\ \beta} \right\}   \nonumber \\
&=& exp\left\{\frac{ s^j_2 q_j^1 }{1 - s^j_1 q_j^1} L_1^{\ 2} \right\} exp\left\{ \log\left( \frac{( s^j_1 q_j^1 - 1)^2}{1 - (s^j_\beta q_j^\beta)  +  s^1_\beta q_1^\beta s^2_\gamma q_2^\gamma - s^1_\beta q_2^\beta s^2_\gamma q_1^\gamma } \right) L_2^{\ 2}  \right\} exp\left\{\frac{s^j_1 q^2_j }{1 - s^j_1 q_j^1} L_2^{\ 1} \right\}  \nonumber
\end{eqnarray}
\begin{eqnarray}
& &  e^{\left( \bar{C}_{\bar{l}}(\bar{s},\bar{q}))^{\dot{\beta}}_{\dot{\alpha}} \bar{L}^{\dot{\alpha}}_{\ \dot{\beta}}  \right) }  \nonumber \\
&=&  exp\left\{ \frac{\bar{s}_{j \dot{1}} \bar{q}^{j \dot{2}} }{1 - \bar{s}_{j \dot{2}} \bar{q}^{j \dot{2}}} \bar{L}^{\dot{1}}_{\ \dot{2}}   \right\} exp\left\{ \log\left(  \frac{ (1 - \bar{s}_{j \dot{2}} \bar{q}^{j \dot{2}})^2}{1 - (\bar{s}_{j \dot{\alpha}} \bar{q}^{j \dot{\alpha}}) +  \bar{s}_{j \dot{1}} \bar{q}^{j \dot{1}} \bar{s}_{k \dot{2}} \bar{q}^{k \dot{2}} -  \bar{s}_{j \dot{2}} \bar{q}^{j \dot{1}} \bar{s}_{k \dot{1}} \bar{q}^{k \dot{2}}}  \right) \bar{L}^{\dot{1}}_{\ \dot{1}}   \right\}\times \nonumber\\
& & exp\left\{ \frac{\bar{s}_{j \dot{2}} \bar{q}^{j \dot{1}} }{1 - \bar{s}_{j \dot{2}} \bar{q}^{j \dot{2}}} \bar{L}^{\dot{2}}_{\ \dot{1}}   \right\}   \nonumber
\end{eqnarray} 
The third swap rule we will use (which may also be derived from (\ref{newBCH})) involves the exponentials for $P$ and $K$.  This rule has been derived and used in \cite{Koch:2021tvp} for general dimensions and provides a good test for the conjecture (\ref{newBCH}).  We restate it here for the four-dimensional example in terms of spinor indices
\begin{equation}
e^{ k_{\alpha \dot{\alpha} } K^{\dot{\alpha} \alpha}  } e^{ p^{\dot{\alpha} \alpha  } P_{\alpha \dot{\alpha}} } = e^{ p'^{\dot{\alpha} \alpha } P_{\alpha \dot{\alpha} } } e^{ d D} e^{\bar{\lambda}_{\dot{1}}^{\dot{2}} \bar{L}^{\dot{1} }_{\ \dot{2}} }   e^{\bar{\lambda}_0 \bar{L}^{\dot{1}}_{\ \dot{1} }  } e^{\bar{\lambda}_{\dot{2}}^{\dot{1}} \bar{L}^{\dot{2} }_{\ \dot{1}} }  e^{\lambda_2^1 L_{1}^{\ 2}} e^{\lambda_0 L_{2}^{\ 2}} e^{\lambda_1^2 L_{2}^{\ 1}} e^{ k'_{\alpha \dot{\alpha} } K^{\dot{\alpha} \alpha} }    \label{KPSwap}
\end{equation}
where 
\begin{eqnarray}
p'^{\dot{\alpha} \alpha} & = &  \frac{4 p^{\dot{\alpha} \beta} k_{\beta \dot{\beta}  }  p^{\dot{\beta} \alpha} - 4 p^{\dot{\beta} \beta} k_{\beta \dot{\beta}} p^{\dot{\alpha} \alpha}  +   p^{\dot{\alpha} \alpha}   }{ 1 - 4 p^{\dot{\beta} \beta} k_{\beta \dot{\beta}}  + 8 (p^{\dot{\beta} \beta} k_{\beta \dot{\beta}})^2 - 8 p^{\dot{\beta} \beta } k_{ \beta \dot{\gamma} }  p^{ \dot{\gamma} \gamma}k_{\gamma \dot{\beta} }  }   \nonumber  \\
d & = & - \log\left(  1 - 4 p^{\dot{\beta} \beta} k_{\beta \dot{\beta}}  + 8 (p^{\dot{\beta} \beta} k_{\beta \dot{\beta}})^2 - 8 p^{\dot{\beta} \beta } k_{ \beta \dot{\gamma} }  p^{ \dot{\gamma} \gamma}k_{\gamma \dot{\beta} }  \right)    \nonumber \\
\lambda_{1}^2 & = & \frac{4 k_{1 \dot{\beta} } p^{\dot{\beta} 2}}{1 - 4 k_{1 \dot{\beta} } p^{\dot{\beta} 1} }    \nonumber \\
\lambda_{2}^1 & = & \frac{4 k_{2 \dot{\beta} } p^{\dot{\beta} 1}}{ 1 - 4 k_{1 \dot{\beta} } p^{\dot{\beta} 1} }   \nonumber \\
\lambda_0 & = & 2 \log\left( \frac{1 - 4 k_{1 \dot{\beta} } p^{\dot{\beta} 1} }{\sqrt{1 - 4 p^{\dot{\beta} \beta} k_{\beta \dot{\beta}}  + 8 (p^{\dot{\beta} \beta} k_{\beta \dot{\beta}})^2 - 8 p^{\dot{\beta} \beta } k_{ \beta \dot{\gamma} }  p^{ \dot{\gamma} \gamma}k_{\gamma \dot{\beta} }}}  \right)    \nonumber \\
\bar{\lambda}^{\dot{1}}_{\dot{2}} & = & \frac{4 p^{\dot{1} \alpha} k_{\alpha \dot{2}}  }{1 - 4 p^{\dot{2} \alpha} k_{\alpha \dot{2}} } \nonumber \\
\bar{\lambda}^{\dot{2}}_{\dot{1}} & = & \frac{4 p^{\dot{2} \alpha} k_{\alpha \dot{1}}  }{1 - 4 p^{\dot{2} \alpha} k_{\alpha \dot{2}} }    \nonumber \\
\bar{\lambda}_0 & = & 2 \log\left(\frac{1 - 4 p^{\dot{2} \alpha} k_{\alpha \dot{2}}   }{   \sqrt{  1 - 4 p^{\dot{\beta} \beta} k_{\beta \dot{\beta}}  + 8 (p^{\dot{\beta} \beta} k_{\beta \dot{\beta}})^2 - 8 p^{\dot{\beta} \beta } k_{ \beta \dot{\gamma} }  p^{ \dot{\gamma} \gamma}k_{\gamma \dot{\beta} } }  }  \right)    \nonumber \\
k'_{ \alpha \dot{\alpha } } & = & \frac{4 k_{ \alpha \dot{\beta}}  p^{\dot{\beta} \beta } k_{ \beta \dot{\alpha}}   -   4  k_{ \beta \dot{\beta}} p^{\dot{\beta} \beta } k_{ \alpha \dot{\alpha}} +  k_{\alpha \dot{\alpha} } }{  1 - 4 p^{\dot{\beta} \beta} k_{\beta \dot{\beta}}  + 8 (p^{\dot{\beta} \beta} k_{\beta \dot{\beta}})^2 - 8 p^{\dot{\beta} \beta } k_{ \beta \dot{\gamma} }  p^{ \dot{\gamma} \gamma}k_{\gamma \dot{\beta} }     }
\end{eqnarray}
Note that the factor 
\begin{eqnarray}
& & 1 - 4 p^{\dot{\beta} \beta} k_{\beta \dot{\beta}}  + 8 (p^{\dot{\beta} \beta} k_{\beta \dot{\beta}})^2 - 8 p^{\dot{\beta} \beta } k_{ \beta \dot{\gamma} }  p^{ \dot{\gamma} \gamma}k_{\gamma \dot{\beta} }  \nonumber \\
&=& (1 - 4 p^{\dot{\beta} 1} k_{1 \dot{\beta}})(1 - 4 p^{\dot{\beta} 2} k_{2 \dot{\beta}}) - 16 p^{\dot{\beta} 2} k_{1 \dot{\beta}} p^{\dot{\beta} 1} k_{2 \dot{\beta}}   \nonumber \\
& = & (1 - 4 p^{\dot{1} \beta} k_{\beta \dot{1}}) (1 - 4 p^{\dot{2} \beta} k_{\beta \dot{2}}) - 16 p^{\dot{2} \beta} k_{\beta \dot{1}} p^{\dot{1} \beta} k_{\beta \dot{2}}   \nonumber
\end{eqnarray}
appears in many of the expressions above.  The first is useful for recasting the variables as traces of matrices while the second and third have contracted the dotted and undotted Greek indices respectively.  Factors with this structure will appear frequently in our expressions and we will change between these equivalent ways of rewriting them for aesthetic reasons.  \\ \\
The mapping from the Lorentz-indexed variables used in \cite{Koch:2021tvp} to the spinor-indexed variables used here is 
\begin{eqnarray}
\left\{ \alpha_1, \alpha_2, \alpha_3, \alpha_4 \right\} & = & \left\{  p^{\dot{1} 1} + p^{\dot{2} 2}  , -i (p^{\dot{1} 2} + p^{\dot{2} 1}), p^{\dot{1} 2} - p^{\dot{2} 1},  -i(p^{\dot{1} 1} - p^{\dot{2} 2})  \right\} \nonumber \\
\left\{ \alpha_1^*, \alpha_2^*, \alpha_3^*, \alpha_4^* \right\} & = & \left\{  k_{ 1 \dot{1}} + k_{ 2 \dot{2}}  , i (k_{ 2 \dot{1}} + k_{ 1 \dot{2}}), k_{ 2 \dot{1}} - k_{ 1 \dot{2}},  i(k_{ 1 \dot{1}} - k_{ 2 \dot{2}})  \right\}
\end{eqnarray}
and, in particular, we have 
\begin{eqnarray}
    1 - 2 \alpha \cdot \alpha^* + (\alpha \cdot \alpha)(\alpha^*\cdot \alpha^*) & = & 1 - 4 p^{\dot{\beta} \beta} k_{\beta \dot{\beta}}  + 8 (p^{\dot{\beta} \beta} k_{\beta \dot{\beta}})^2 - 8 p^{\dot{\beta} \beta } k_{ \beta \dot{\gamma} }  p^{ \dot{\gamma} \gamma}k_{\gamma \dot{\beta} }   \nonumber \\
    \delta_\alpha^{\beta} - \alpha^*_\mu \alpha_\nu (\sigma^\mu)_{\alpha \dot{\beta}}(\bar{\sigma}^\nu)^{\dot{\beta} \beta} & = & (1 - 4 k_{2\dot{\beta}}p^{\dot{\beta} 2 } ) \delta^1_\alpha \delta^\beta_1  +   (1 - 4 k_{1\dot{\beta}}p^{\dot{\beta} 1 } ) \delta^2_\alpha \delta^\beta_2   \nonumber \\
    & & + 4 k_{2 \dot{\beta}} p^{\dot{\beta} 1} \delta_\alpha^2 \delta^\beta_1 + 4 k_{1 \dot{\beta}} p^{\dot{\beta} 2} \delta_\alpha^1 \delta^\beta_2    \label{twoPtvars}
\end{eqnarray}
In our computations we supplement the swap rules (\ref{SQSwap}), (\ref{SbQbSwap}), (\ref{KPSwap}) with the well-known formulas
\begin{eqnarray}
e^{A} e^{B} & = & e^{B + [A, B]} e^{A} \ \ \ \textnormal{if} \ \ \ [A, [A, B]] = 0     \nonumber \\
e^{A} e^{B} & = & e^{e^k B} e^{A} \ \  \ \ \ \ \ \textnormal{if} \ \ \ [A, B]=kB      \nonumber
\end{eqnarray}
as well as decomposition formulas for $SU(2)$.  

\section{Results}
\label{sectionResults}

We are now in a position to compute the overlaps and resulting super-K\"{a}hler potential.  As detailed in appendix B significant pieces of the state overlap can be computed without needing to specify the value for $\mathcal{N}$.  Specifically, the parts depending on the spins $h, \bar{h}$ can be extracted in full detail.  The intermediate result is stated in (\ref{Master2}). In order to proceed from this point we need to specify the transformation properties of the reference state under the R-charge transformations.  This allows us to compute the $\Delta$ and $R$-charge dependent expectation value in (\ref{Master2}).  Note that this expectation value is taken w.r.t. the state $|\psi'\rangle$ which carries the same scaling dimension and $R$-charge as $|\psi_0\rangle$ but is spinless.  

\subsection{No supersymmetry ($\mathcal{N} =0$)}

We begin with the case where there is no supersymmetry.   This is a known result from \cite{Koch:2021tvp} and serves as a good check that our expressions are correct.  When the reference state is the highest weight state $\psi_0 = |\Delta ; h,h ; \bar{h}, \bar{h}\rangle$ we may proceed directly from (\ref{Master2}) for which 
\begin{eqnarray}
& & \langle   \psi' | e^{\bar{r}_i^j R^i_j} e^{ \bar{C}_{d}( \bar{s}, 2 p\cdot s ) D  } e^{ \bar{C}_{r}( \bar{s}, 2 p\cdot s )^j_i R_j^i  } e^{\bar{s}_{i \dot{\alpha}}'' \bar{S}^{i \dot{\alpha}}  }   e^{(C_r(s'' , q''  ))_i^j R^i_j }  e^{ C_d(s'', q''  ) D}  e^{   (\bar{q}'')^{i \dot{\alpha} } \bar{Q}_{i \dot{\alpha} }  }  e^{ \bar{C}_{r}( 2 q \cdot k, \bar{q} )^j_i R_j^i  } e^{ \bar{C}_{d}( 2 q \cdot k, \bar{q} )  D  } e^{r_i^j R^i_j}| \psi' \rangle     \nonumber \\
&\rightarrow & 1   \nonumber
\end{eqnarray}
This substitution as well as  dropping the supercharge dependence, yields
\begin{eqnarray}
&&  \langle \psi_{0} | e^{ \bar{l}_{\dot{2}}^{\dot{1}}  \bar{L}^{\dot{2}}_{\ \dot{1}} }  e^{ l^{2}_1  L_{2}^{\ 1}} e^{ k_{ \alpha  \dot{\alpha}} K^{\dot{\alpha} \alpha}  } e^{ p^{ \dot{\alpha} \alpha } P_{\alpha \dot{\alpha}} } e^{ l^{1}_2  L_{1}^{\ 2}} e^{ \bar{l}_{\dot{1}}^{\dot{2}} \bar{L}^{\dot{1}}_{\ \dot{2}} }  |\psi_{0}\rangle     \nonumber \\
&=& \left( (1 - 4 k_{1 \dot{\beta} } p^{\dot{\beta} 1}) (1 - 4 k_{2 \dot{\beta} } p^{\dot{\beta} 2})  - 16 k_{1 \dot{\beta} } p^{\dot{\beta} 2} k_{2 \dot{\gamma} } p^{\dot{\gamma} 1} \right)^{-\Delta} \times \nonumber \\
& &  \left(   \frac{(1 - 4 k_{1 \dot{\beta} } p^{\dot{\beta} 1}) + 4 l^2_1 k_{2 \dot{\beta} } p^{\dot{\beta} 1}    +   4 l^1_2  k_{1 \dot{\beta} } p^{\dot{\beta} 2}  +   l^2_1 l^1_2 (1 - 4 k_{2 \dot{\beta} } p^{\dot{\beta} 2})       }{\sqrt{   (1 - 4 k_{1 \dot{\beta} } p^{\dot{\beta} 1}) (1 - 4 k_{2 \dot{\beta} } p^{\dot{\beta} 2})  - 16 k_{1 \dot{\beta} } p^{\dot{\beta} 2} k_{2 \dot{\gamma} } p^{\dot{\gamma} 1}    }  }      \right)^{2 h}  \times \nonumber   \\
& &  \left(   \frac{(1 - 4  p^{\dot{2} \beta} k_{\beta \dot{2}}   ) + 4  \bar{l}^{\dot{1}}_{
\dot{2}} p^{\dot{2} \beta} k_{\beta \dot{1}}    +   4 \bar{l}^{\dot{2}}_{
\dot{1}}  p^{\dot{1} \beta} k_{\beta \dot{2}}  +    \bar{l}^{\dot{1}}_{
\dot{2}}  \bar{l}^{\dot{2}}_{
\dot{1}} (1 - 4 p^{\dot{1} \beta} k_{\beta \dot{1}})       }{   \sqrt{(1 - 4  p^{\dot{2} \beta} k_{\beta \dot{2}}   ) (1 - 4 p^{\dot{1} \beta} k_{\beta \dot{1}})   - 16 p^{\dot{1} \beta} k_{\beta \dot{2}} p^{\dot{2} \beta} k_{\beta \dot{1}}      }   }      \right)^{2 \bar{h}}     \nonumber \\
&=& \left( (1 - 4 k_{1 \dot{\beta} } p^{\dot{\beta} 1}) (1 - 4 k_{2 \dot{\beta} } p^{\dot{\beta} 2})  - 16 k_{1 \dot{\beta} } p^{\dot{\beta} 2} k_{2 \dot{\gamma} } p^{\dot{\gamma} 1} \right)^{-(\Delta + h + \bar{h}) } \times \nonumber \\
& &  \left( (1 - 4 k_{1 \dot{\beta} } p^{\dot{\beta} 1}) + 4 l^2_1 k_{2 \dot{\beta} } p^{\dot{\beta} 1}    +   4 l^1_2  k_{1 \dot{\beta} } p^{\dot{\beta} 2}  +   l^2_1 l^1_2 (1 - 4 k_{2 \dot{\beta} } p^{\dot{\beta} 2})            \right)^{2 h}  \times \nonumber   \\
& &  \left(  (1 - 4  p^{\dot{2} \beta} k_{\beta \dot{2}}   ) + 4  \bar{l}^{\dot{1}}_{
\dot{2}} p^{\dot{2} \beta} k_{\beta \dot{1}}    +   4 \bar{l}^{\dot{2}}_{
\dot{1}}  p^{\dot{1} \beta} k_{\beta \dot{2}}  +    \bar{l}^{\dot{1}}_{
\dot{2}}  \bar{l}^{\dot{2}}_{
\dot{1}} (1 - 4 p^{\dot{1} \beta} k_{\beta \dot{1}})      \right)^{2 \bar{h}}  \label{Nis0Overlap}
\end{eqnarray}
This expression matches precisely the one obtained in \cite{Koch:2021tvp} after mapping the Lorentz-indexed variables to the spinor-indexed variables.  In the spinor basis a nice structure of the overlap is apparent - all the dotted and undotted indices appear in pairs e.g. the number of $
\dot{1}$ subscript indices is equal to the number of $\dot{1}$ superscript indices.  We also note that the diagonal entries of $p^{\dot{\alpha} \gamma} k_{\gamma \dot{\beta}} $ and  $ k_{\beta \dot{\gamma}} p^{\dot{\gamma} \alpha} $ appear with an additional factor as compared to the off-diagonal entries.   \\ \\ 
The K\"{a}hler potential is given by the logarithm of the overlap.  Where the overlap has product structure with powers involving the scaling dimension and spin, the K\"{a}hler potential has a sum structure with the coefficient in front of the terms featuring the scaling dimension and spin.  

\subsection{$\mathcal{N} = 1$ supersymmetry}

For $\mathcal{N}=1$ there is a single $R$-charge, namely $R_1^1$.  In our expressions we also suppress the Latin index, since this always assumes the value of $i=1$.   The reference state may be chosen so that, in addition to the conditions (\ref{refState}), it satisfies
\begin{equation}
R_1^1 |\psi_0\rangle = R |\psi_0\rangle 
\end{equation}
The generator $R_1^1$ forms part of the stability subgroup.  Explicitly, the reference state is given by
\begin{equation}
|\psi_0\rangle = |\Delta; h, h; \bar{h}, \bar{h}; R\rangle
\end{equation}
  The coefficients appearing in the expansion formulas (\ref{SQSwap}) and (\ref{SbQbSwap}) may be written in the compact form
\begin{eqnarray}
(C_q(s, q))^\alpha Q_\alpha & = & (1 - s\cdot q) q \cdot Q   \nonumber \\
C_r(s, q)  & = & \log\left( (1 - s_1 q^1)(1 - s_2 q^2) - s_1 q^2 s_2 q^1  \right)  = -\log\left( 1 + s\cdot q \right) \nonumber \\
C_d(s, q) & = & - \frac{1}{2} C_r(s, q)  \nonumber \\
(C_s(s, q))_\alpha S^\alpha & = & (1 - s\cdot q) s \cdot S  \nonumber  \\ 
 \\
(\bar{C}_{\bar{q}}(\bar{s}, \bar{q}) )^{\dot{\alpha}} \bar{Q}_{\dot{\alpha} } & = & (1 - \bar{s} \cdot \bar{q}) \bar{q} \cdot \bar{Q}  \nonumber \\
\bar{C}_r(\bar{s}, \bar{q}) & = &  -\log\left( (1 - \bar{s}_{\dot{1}} \bar{q}^{\dot{1}  }  )(1 - \bar{s}_{\dot{2}}  \bar{q}^{\dot{2} }  ) - \bar{s}_{\dot{1} } \bar{q}^{\dot{2}} \bar{s}_{\dot{2} } \bar{q}^{\dot{1} }  \right)  = \log\left( 1 + \bar{s}\cdot \bar{q}   \right) \nonumber   \\
\bar{C}_d & = & \frac{1}{2} \bar{C}_r(\bar{s}, \bar{q}) \nonumber \\
(\bar{C}_{\bar{s}}(\bar{s}, \bar{q}) )_{\dot{\alpha}} \bar{S}^{\dot{\alpha} } & = & (1 - \bar{s}\cdot \bar{q} )\bar{s} \cdot \bar{S}     \nonumber
\end{eqnarray}
The expressions above should be thought of as expanded to first order in the Grassmann variables.  We find that the functions quoted above give rise to more compact expressions, however.  \\ \\
To compute the super-K\"{a}hler potential we need to follow on from the general expression (\ref{Master2}).  For $\mathcal{N}=1$ this gives
\begin{eqnarray}
& & \langle   \psi' | e^{\bar{r}_i^j R^i_j} e^{ \bar{C}_{d}( \bar{s}, 2 p\cdot s ) D  } e^{ \bar{C}_{r}( \bar{s}, 2 p\cdot s )^j_i R_j^i  } e^{\bar{s}_{i \dot{\alpha}}'' \bar{S}^{i \dot{\alpha}}  }   e^{(C_r(s'' , q''  ))_i^j R^i_j }  e^{ C_d(s'', q''  ) D}  e^{   (\bar{q}'')^{i \dot{\alpha} } \bar{Q}_{i \dot{\alpha} }  }  e^{ \bar{C}_{r}( 2 q \cdot k, \bar{q} )^j_i R_j^i  } e^{ \bar{C}_{d}( 2 q \cdot k, \bar{q} )  D  } e^{r_i^j R^i_j}| \psi' \rangle     \nonumber \\
&\rightarrow & \langle \psi'| e^{\bar{s}_{i \dot{\alpha}}'' \bar{S}^{i \dot{\alpha}}  }   e^{C_r(s'' , q''  ) R^1_1 }   e^{  e^{\frac{C_d(s'', q''  )}{2} } (\bar{q}'')^{i \dot{\alpha} } \bar{Q}_{i \dot{\alpha} }  }       | \psi'\rangle \ e^{ \left(   \bar{C}_{r}( 2 q \cdot k, \bar{q} )  +   \bar{C}_{r}( \bar{s}, 2 p\cdot s )  \right) (R + \frac{\Delta}{2}) } e^{ C_d(s'' , q'') \Delta } \nonumber  \\
& = & \langle \psi'| e^{\bar{s}_{ \dot{\alpha}}'' \bar{S}^{\dot{\alpha}}  }  e^{  e^{\frac{1}{2}C_d(s'', q''  )   -  \frac{3}{4} C_r(s'' , q''  ) } (\bar{q}'')^{ \dot{\alpha} } \bar{Q}_{ \dot{\alpha} }  }       | \psi'\rangle \ e^{ \left(   \bar{C}_{r}( 2 q \cdot k, \bar{q} )  +   \bar{C}_{r}( \bar{s}, 2 p\cdot s ) +   C_r(s'' , q''  )  \right) (R + \frac{\Delta}{2}) }  e^{ C_d(s'' , q'') \Delta }   \nonumber \\
& = & \langle \psi'| e^{\bar{s}_{ \dot{\alpha}}'' \bar{S}^{ \dot{\alpha}}  }  e^{  (1 + s'' \cdot q'') (\bar{q}'')^{ \dot{\alpha} } \bar{Q}_{ \dot{\alpha} }  }       | \psi'\rangle \ e^{ \left(   \bar{C}_{r}( 2 q \cdot k, \bar{q} )  +   \bar{C}_{r}( \bar{s}, 2 p\cdot s )   \right) (R + \frac{\Delta}{2}) } e^{ C_r(s'' , q''  ) (R - \frac{\Delta}{2} ) }    \nonumber \\
& = & e^{ \left(   \bar{C}_{r}( 2 q \cdot k, \bar{q} )  +   \bar{C}_{r}( \bar{s}, 2 p\cdot s ) +    \bar{C}_{r}(\bar{s}'',  (1 + s''\cdot q'') \bar{q}'')  \right) (R + \frac{\Delta}{2}) } e^{ \left(C_r(s'' , q'')  \right)(R - \frac{\Delta}{2})     }      \nonumber \\
&=& \left( 1 + 2 q^\alpha k_{\alpha \dot{\alpha}} \bar{q}^{\dot{\alpha}}   \right)^{\frac{\Delta}{2} + R} \left( 1 + 2 \bar{s}_{\dot{\alpha} } p^{\dot{\alpha} \alpha} s_{   \alpha }   \right)^{\frac{\Delta}{2} + R}  \left( 1 + (1 + s''\cdot q'')\bar{s}'' \cdot \bar{q}''  \right)^{\frac{\Delta}{2} + R} \left(1 + s'' \cdot q''\right)^{\frac{\Delta}{2} - R  }    \nonumber \\
&=& \left( 1 + 2 q^\alpha k_{\alpha \dot{\alpha}} \bar{q}^{\dot{\alpha}}   \right)^{\frac{\Delta}{2} + R} \left( 1 + 2 \bar{s}_{\dot{\alpha} } p^{\dot{\alpha} \alpha} s_{   \alpha }   \right)^{\frac{\Delta}{2} + R}  \left( \frac{1}{(1 + s''\cdot q'')} + \bar{s}'' \cdot \bar{q}''  \right)^{\frac{\Delta}{2} + R} \left(1 + s'' \cdot q''\right)^{ \Delta  }  
\end{eqnarray}  
After simplifying this result a bit we find the expectation value
\begin{eqnarray}
& & \langle \psi_{0} | e^{ \bar{l}_{\dot{2}}^{\dot{1}} \bar{L}^{\dot{2}}_{\ \dot{1}} }  e^{ l^{2}_1 L_{2}^{\ 1}} e^{ \bar{s}_{i \dot{\alpha}} \bar{S}^{i \dot{\alpha}}  } e^{ s_\alpha^i S^\alpha_i }   e^{ k_{\alpha \dot{\alpha}} K^{\dot{\alpha} \alpha } }   e^{ p^{ \dot{\alpha} \alpha} P_{ \alpha \dot{\alpha}} } e^{q^\alpha_i Q^i_\alpha} e^{\bar{q}^{i \dot{\alpha}} \bar{Q}_{i \dot{\alpha}}  } e^{ l^{1}_2 L_{1}^{\ 2}} e^{ \bar{l}_{\dot{1}}^{\dot{2}} \bar{L}^{\dot{1}}_{\ \dot{2}} }  |\psi_{0}\rangle   \nonumber \\
&=& \left( (1 - 4 k_{1 \dot{\beta} } p^{\dot{\beta} 1} - s_1 q^1) (1 - 4 k_{2 \dot{\beta} } p^{\dot{\beta} 2} - s_2 q^2 )  - ( 4 k_{1 \dot{\beta} } p^{\dot{\beta} 2} + s_1 q^2)( 4 k_{2 \dot{\beta} } p^{\dot{\beta} 1} + s_2 q^1 ) \right)^{-\Delta}  \times \nonumber \\
& &  \left(   \frac{(1 - 4 k_{1 \dot{\beta} } p^{\dot{\beta} 1} - s_1 q^1) +  l^2_1 (4k_{2 \dot{\beta} } p^{\dot{\beta} 1}  +  s_2 q^1)  +   l^1_2 ( 4 k_{1 \dot{\beta} } p^{\dot{\beta} 2}  +  s_1 q^2  ) +    l^2_1  l^1_2 (1 - 4 k_{2 \dot{\beta} } p^{\dot{\beta} 2} - s_2 q^2)       }{\sqrt{   (1 - 4 k_{1 \dot{\beta} } p^{\dot{\beta} 1}   -  s_1 q^1) (1 - 4 k_{2 \dot{\beta} } p^{\dot{\beta} 2} - s_2 q^2)  - (4 k_{1 \dot{\beta} } p^{\dot{\beta} 2} +   s_1 q^2  )(4 k_{2 \dot{\beta} } p^{\dot{\beta} 1} +   s_2 q^1   )   }  }      \right)^{2 h }  \times \nonumber   \\
& &  \left(   \frac{(1 - 4  \tilde{p}^{\dot{2} \beta} \tilde{k}_{\beta \dot{2}} - \bar{s}_{ \dot{2}}\bar{q}^{ \dot{2}}   ) +  \bar{l}^{\dot{1}}_{
\dot{2}} (   4 \tilde{p}^{\dot{2} \beta} \tilde{k}_{\beta \dot{1}} +  \bar{s}_{ \dot{1}}\bar{q}^{ \dot{2}}  )   +   \bar{l}^{\dot{2}}_{
\dot{1}} (   4 \tilde{p}^{\dot{1} \beta} \tilde{k}_{\beta \dot{2}} +   \bar{s}_{ \dot{2}} \bar{q}^{ \dot{1}}   ) +    \bar{l}^{\dot{1}}_{
\dot{2}}  \bar{l}^{\dot{2}}_{
\dot{1}} (1 - 4 \tilde{p}^{\dot{1} \beta} \tilde{k}_{\beta \dot{1}} - \bar{s}_{ \dot{1}}\bar{q}^{ \dot{1}})       }{   \sqrt{(1 - 4  \tilde{p}^{\dot{1} \beta} \tilde{k}_{\beta \dot{1}} - \bar{s}_{ \dot{1}}\bar{q}^{ \dot{1}}) (1 - 4  \tilde{p}^{\dot{2} \beta} \tilde{k}_{\beta \dot{2}} - \bar{s}_{ \dot{2}}\bar{q}^{ \dot{2}})  - ( 4 \tilde{p}^{\dot{2} \beta} \tilde{k}_{\beta \dot{1}}  + \bar{s}_{ \dot{1}}\bar{q}^{ \dot{2}}) (4  \tilde{p}^{\dot{1} \beta} \tilde{k}_{\beta \dot{2}}  +   \bar{s}_{ \dot{2}}\bar{q}^{ \dot{1}}   )    }   }      \right)^{2 \bar{h}}  \times  \nonumber \\
  & &\left( \frac{\left( (1 - 4 k_{1 \dot{\beta} } p^{\dot{\beta} 1} - s_1 q^1) (1 - 4 k_{2 \dot{\beta} } p^{\dot{\beta} 2} - s_2 q^2 )  - ( 4 k_{1 \dot{\beta} } p^{\dot{\beta} 2} + s_1 q^2)( 4 k_{2 \dot{\beta} } p^{\dot{\beta} 1} + s_2 q^1 ) \right)   }{\left( (1 - 4 k_{1 \dot{\beta} } p^{\dot{\beta} 1}) (1 - 4 k_{2 \dot{\beta} } p^{\dot{\beta} 2} )  - ( 4 k_{1 \dot{\beta} } p^{\dot{\beta} 2} )( 4 k_{2 \dot{\beta} } p^{\dot{\beta} 1}) \right)(1 + 2 q^{\alpha} k_{\alpha \dot{\alpha}} \bar{q}^{\dot{\alpha}}  )^{-1}(1 + 2 \bar{s}_{\dot{\alpha}} p^{\dot{\alpha} \alpha } s_{\alpha}  )^{-1} }   \right. \nonumber \\
  & & \left. + \frac{(1 - 4  p^{\dot{1} \beta} k_{\beta \dot{1}}) (1 - 4 p^{\dot{2} \beta} k_{\beta \dot{2}} )  - ( 4 p^{\dot{2} \beta} k_{\beta \dot{1}}) (4  p^{\dot{1} \beta} k_{\beta \dot{2}}  ) }{(1 - 4  p^{\dot{1} \beta} k_{\beta \dot{1}} - \bar{s}_{\dot{1}} \bar{q}^{\dot{1}}) (1 - 4 p^{\dot{2} \beta} k_{\beta \dot{2}} - \bar{s}_{\dot{2}}\bar{q}^{\dot{2}})  - ( 4 p^{\dot{2} \beta} k_{\beta \dot{1}} + \bar{s}_{\dot{1}} \bar{q}^{\dot{2}}) (4  p^{\dot{1} \beta} k_{\beta \dot{2}} + \bar{s}_{\dot{2}} \bar{q}^{\dot{1}} )} -1 \right)^{\frac{\Delta}{2} + R}     \label{Nis1Overlap}
\end{eqnarray}
where 
\begin{eqnarray}
\tilde{k}_{ \alpha \dot{\alpha }} & = & k_{ \alpha \dot{\alpha}} - \frac{1}{2}  s_{\alpha} \bar{s}_{\dot{\alpha}}   \nonumber \\
\tilde{p}^{ \dot{\alpha } \alpha } & = & p^{ \dot{\alpha } \alpha } - \frac{1}{2} \bar{q}^{\dot{\alpha}} q^{\alpha} 
\end{eqnarray}
Due to the Grassmann variables there are many equivalent ways to write (\ref{Nis1Overlap}).  When compared to (\ref{Nis0Overlap}) we note that, in the spinor index notation, the same pattern observed therein persists-   superscript and subscript indices, both the Greek, dotted Greek and Latin indices are balanced.  If we set the Grassman indices to zero we recover  (\ref{Nis0Overlap}), as expected.  \\ \\ 
Note that, unlike in (\ref{Nis0Overlap}), the denominator in the $h, \bar{h}$ dependent pieces is different from the $\Delta$-dependent contribution on the first line due to the presence of the Grassmann numbers.  The super-K\"{a}hler potential is given by the logarithm of the above overlap.  As was the case for the non-supersymmetric potential the structure is a sum of terms with the coefficients in front featuring the various quantum numbers labelling the reference state.  \\ \\
An instructive limit of the overlap (\ref{Nis1Overlap}) is to consider the case $\bar{q}^{i\dot{\alpha}} \rightarrow 0$, $\bar{s}_{i\dot{\alpha}} \rightarrow 0$.  In this case the overlap becomes 
\begin{eqnarray}
& & \langle \psi_{0} | e^{ \bar{l}_{\dot{2}}^{\dot{1}} \bar{L}^{\dot{2}}_{\ \dot{1}} }  e^{ l^{2}_1 L_{2}^{\ 1}} e^{ s_\alpha^i S^\alpha_i }   e^{ k_{\alpha \dot{\alpha}} K^{\dot{\alpha} \alpha } }   e^{ p^{ \dot{\alpha} \alpha} P_{ \alpha \dot{\alpha}} } e^{q^\alpha_i Q^i_\alpha}  e^{ l^{1}_2 L_{1}^{\ 2}} e^{ \bar{l}_{\dot{1}}^{\dot{2}} \bar{L}^{\dot{1}}_{\ \dot{2}} }  |\psi_{0}\rangle    \nonumber \\
&=&  \left((1 - 4 k_{1 \dot{\beta} } p^{\dot{\beta} 1} - s_1 q^1) +  l^2_1 (4k_{2 \dot{\beta} } p^{\dot{\beta} 1}  +  s_2 q^1)  +   l^1_2 ( 4 k_{1 \dot{\beta} } p^{\dot{\beta} 2}  +  s_1 q^2  ) +    l^2_1  l^1_2 (1 - 4 k_{2 \dot{\beta} } p^{\dot{\beta} 2} - s_2 q^2)  \right)^h   \times \nonumber \\
& & \left( (1 - 4  \tilde{p}^{\dot{2} \beta} \tilde{k}_{\beta \dot{2}}  ) +  \bar{l}^{\dot{1}}_{
\dot{2}} (   4 \tilde{p}^{\dot{2} \beta} \tilde{k}_{\beta \dot{1}}  )   +   \bar{l}^{\dot{2}}_{
\dot{1}} (   4 \tilde{p}^{\dot{1} \beta} \tilde{k}_{\beta \dot{2}}   ) +    \bar{l}^{\dot{1}}_{
\dot{2}}  \bar{l}^{\dot{2}}_{
\dot{1}} (1 - 4 \tilde{p}^{\dot{1} \beta} \tilde{k}_{\beta \dot{1}})    \right)^{\bar{h}}  \times \nonumber \\
& &  \left( (1 - 4 k_{1 \dot{\beta} } p^{\dot{\beta} 1} - s_1 q^1) (1 - 4 k_{2 \dot{\beta} } p^{\dot{\beta} 2} - s_2 q^2 )  - ( 4 k_{1 \dot{\beta} } p^{\dot{\beta} 2} + s_1 q^2)( 4 k_{2 \dot{\beta} } p^{\dot{\beta} 1} + s_2 q^1 ) \right)^{-\frac{\Delta}{2} + R - h} \times  \nonumber \\
& & \left( (1 - 4 k_{1 \dot{\beta} } p^{\dot{\beta} 1}) (1 - 4 k_{2 \dot{\beta} } p^{\dot{\beta} 2} )  - ( 4 k_{1 \dot{\beta} } p^{\dot{\beta} 2} )( 4 k_{2 \dot{\beta} } p^{\dot{\beta} 1}) \right)^{-\frac{\Delta}{2} - R - \bar{h}}   \nonumber \\
&=&  \left((1 - 4 k_{1 \dot{\beta} } p^{\dot{\beta} 1} - s_1 q^1) +  l^2_1 (4k_{2 \dot{\beta} } p^{\dot{\beta} 1}  +  s_2 q^1)  +   l^1_2 ( 4 k_{1 \dot{\beta} } p^{\dot{\beta} 2}  +  s_1 q^2  ) +    l^2_1  l^1_2 (1 - 4 k_{2 \dot{\beta} } p^{\dot{\beta} 2} - s_2 q^2)  \right)^h   \times \nonumber \\
& & \left( (1 - 4  \tilde{p}^{\dot{2} \beta} \tilde{k}_{\beta \dot{2}}  ) +  \bar{l}^{\dot{1}}_{
\dot{2}} (   4 \tilde{p}^{\dot{2} \beta} \tilde{k}_{\beta \dot{1}}  )   +   \bar{l}^{\dot{2}}_{
\dot{1}} (   4 \tilde{p}^{\dot{1} \beta} \tilde{k}_{\beta \dot{2}}   ) +    \bar{l}^{\dot{1}}_{
\dot{2}}  \bar{l}^{\dot{2}}_{
\dot{1}} (1 - 4 \tilde{p}^{\dot{1} \beta} \tilde{k}_{\beta \dot{1}})    \right)^{\bar{h}}  \times \nonumber \\
& & \left( \frac{ (1 - 4 k_{1 \dot{\beta} } p^{\dot{\beta} 1}) (1 - 4 k_{2 \dot{\beta} } p^{\dot{\beta} 2} )  - ( 4 k_{1 \dot{\beta} } p^{\dot{\beta} 2} )( 4 k_{2 \dot{\beta} } p^{\dot{\beta} 1})   + 2 (\frac{\Delta}{2} - R + h -1)(\frac{\Delta}{2} - R + h) s_1 s_2 q^2 q^1 }{ \left( (1 - 4 k_{1 \dot{\beta} } p^{\dot{\beta} 1}) (1 - 4 k_{2 \dot{\beta} } p^{\dot{\beta} 2} )  - ( 4 k_{1 \dot{\beta} } p^{\dot{\beta} 2} )( 4 k_{2 \dot{\beta} } p^{\dot{\beta} 1})   \right)^{\Delta + h + \bar{h} + 1}}       \right. \nonumber \\
& & \left. \ \ \ + \ (\frac{\Delta}{2} + h - R)\frac{(1 - 4 k_{2 \dot{\beta}} p^{\dot{\beta} 2} ) s_1 q^1 + 4 k_{2 \dot{\beta}} p^{\dot{\beta 1}} s_1 q^2 + 4 k_{1 \dot{\beta}} p^{\dot{\beta 2}} s_2 q^1 +  (1 - 4 k_{1 \dot{\beta}} p^{\dot{\beta} 1} ) s_2 q^2 }{\left( (1 - 4 k_{1 \dot{\beta} } p^{\dot{\beta} 1}) (1 - 4 k_{2 \dot{\beta} } p^{\dot{\beta} 2} )  - ( 4 k_{1 \dot{\beta} } p^{\dot{\beta} 2} )( 4 k_{2 \dot{\beta} } p^{\dot{\beta} 1})   \right)^{\Delta + h + \bar{h} + 1}}   \right)
\end{eqnarray}
Though we have not unpacked the first factor in the last equality, the structure of the third factor is appealing.  The factors we observe are precisely those in (\ref{twoPtvars}), related to the scalar and fermion two-point functions respectively.  The zero'th order and second order terms in $q$ are bosonic, with the scaling dimension differing by 1.  The first order term in $q$ is fermionic with the appropriate scaling dimension.  \\ \\
When considering the limit $q^\alpha_i \rightarrow 0$, $s_\alpha^i \rightarrow 0$ we find a very similar structure for the overlap in the barred Grassmann variables.

\subsection{$\mathcal{N}=2$ supersymmetry}

In the case $\mathcal{N}=2$ the R-charges form a $U(2)$ \cite{Eberhardt:2020cxo}.  The combination $r_I = \frac{1}{2}\left( R_1^1 + R_2^2\right)$ commutes with all other combinations of $R$-charges.  The reference state may be chosen as a simultaneous eigenvector of $r_I$ and $R_3 = \frac{1}{2}\left( R_2^2 - R_1^1 \right)$ and, in particular, it may be chosen to be a highest weight state
\begin{eqnarray}
r_I |\psi_0\rangle & = & r |\psi_0\rangle \nonumber \\
R_3 |\psi_0\rangle & = & R |\psi_0\rangle \nonumber \\
R^2_1 |\psi_0\rangle & = & 0 \nonumber
\end{eqnarray}
The stability subgroup of the reference state is thus spanned by $r_I, R_3$ and $R^2_1$ so that 
\begin{equation}
N e^{r_i^j R^i_j} |\psi_{0}\rangle = e^{r_1^2 R^1_2} |\psi_0\rangle 
\end{equation}
Explicitly, the reference state is thus 
\begin{equation}
|\psi_0\rangle = |\Delta; h, h; \bar{h}, \bar{h}; R, R; r\rangle
\end{equation}
The $R$-charge exponentials in (\ref{SQSwap}) and (\ref{SbQbSwap}) may be further decomposed as 
\begin{eqnarray}
e^{   (C_{r}(s,q))^j_i R_j^i} & = & e^{ -\frac{s^2_\beta q^\beta_1 }{1 + s^1_\beta q^\beta_1} R^1_2   }  e^{\log\left( \frac{(1 +  s^1_\beta q^\beta_1 )^2}{1 + s^i_\alpha q^\alpha_i + \frac{1}{2}(s^i_\alpha q^\alpha_i)^2 +  \frac{1}{2}s^i_\alpha q^\beta_i s^j_\beta q^\alpha_j}  \right) \frac{R_2^2 - R_1^1}{2}  }  e^{ -\frac{s^1_\beta q^\beta_2 }{1 + s^1_\beta q^\beta_1} R^2_1   }   \times   \nonumber \\
& &   e^{\log\left( 1 - s^i_\alpha q^\alpha_i + \frac{1}{2}(s^i_\alpha q^\alpha_i)^2 - \frac{1}{2} s^i_\alpha q^\beta_i s^j_\beta q^\alpha_j     \right) \frac{R_1^1 + R_2^2}{2}   }     \nonumber \\
e^{(\bar{C}_{r}(\bar{s},\bar{q}))^j_i R_j^i} & = & e^{ \frac{\bar{s}_{1 \dot{\alpha}}   \bar{q}^{2 \dot{\alpha} }   }{1 + \bar{s}_{2 \dot{\alpha}}   \bar{q}^{2 \dot{\alpha} }   } R^1_2   } e^{ \log\left( \frac{(1 + \bar{s}_{2 \dot{\alpha}}  \bar{q}^{2 \dot{\alpha} })^2   }{ 1 +   \bar{s}_{i \dot{\alpha} } \bar{q}^{i \dot{\alpha}} +   \frac{1}{2} (\bar{s}_{i \dot{\alpha} } \bar{q}^{i \dot{\alpha}})^2 +   \frac{1}{2}\bar{s}_{i \dot{\alpha} } \bar{q}^{i \dot{\beta}} \bar{s}_{j \dot{\beta} } \bar{q}^{j \dot{\alpha} }    }  \right) \frac{R^2_2 - R^1_1}{2}   }   e^{ \frac{\bar{s}_{2 \dot{\alpha}}\bar{q}^{1 \dot{\alpha} }   }{1 + \bar{s}_{2 \dot{\alpha}}\bar{q}^{2 \dot{\alpha} }   } R^2_1   }      \times    \nonumber \\
& & e^{- \log\left( 1 - \bar{s}_{i \dot{\alpha} } \bar{q}^{i \dot{\alpha}} +   \frac{1}{2} (\bar{s}_{i \dot{\alpha} } \bar{q}^{i \dot{\alpha}})^2 -   \frac{1}{2}\bar{s}_{i \dot{\alpha} } \bar{q}^{i \dot{\beta}} \bar{s}_{j \dot{\beta} } \bar{q}^{j \dot{\alpha} }     \right)\frac{R_1^1 + R_2^2}{2}}
\end{eqnarray}
This implies 
\begin{eqnarray}
e^{(\bar{C}_{r}(\bar{s},2 p\cdot s))^j_i R_j^i} & = & e^{ \frac{2 \bar{s}_{1 \dot{\alpha}} p^{\dot{\alpha} \alpha}  \bar{s}^{2}_\alpha }{1 + 2\bar{s}_{2 \dot{\alpha}} p^{\dot{\alpha} \alpha }  s^{2 }_{\alpha}   } R^1_2   } e^{ \log\left( \frac{(1 + 2\bar{s}_{2 \dot{\alpha}} p^{\dot{\alpha} \alpha }  s^{2 }_{\alpha} )^2   }{ 1 +   2\bar{s}_{i \dot{\alpha} } p^{\dot{\alpha} \alpha} q^i_{\alpha}  +  2 (\bar{s}_{i \dot{\alpha} } p^{\dot{\alpha} \alpha} q^i_{\alpha})^2 +   2 \bar{s}_{i \dot{\alpha} } p^{\dot{\beta} \alpha} s_\alpha^i  \bar{s}_{j \dot{\beta} } p^{\dot{\alpha} \beta} s^j_\beta     }  \right) \frac{R^2_2 - R^1_1}{2}   }   e^{ \frac{2 \bar{s}_{1 \dot{\alpha}} p^{\dot{\alpha} \alpha}  \bar{s}^{2}_\alpha }{1 + 2\bar{s}_{2 \dot{\alpha}} p^{\dot{\alpha} \alpha }  s^{2 }_{\alpha}   } R^2_1   }      \times    \nonumber \\
& & e^{- \log\left( 1 -   2\bar{s}_{i \dot{\alpha} } p^{\dot{\alpha} \alpha} q^i_{\alpha}  +  2 (\bar{s}_{i \dot{\alpha} } p^{\dot{\alpha} \alpha} q^i_{\alpha})^2 -   2 \bar{s}_{i \dot{\alpha} } p^{\dot{\beta} \alpha} s_\alpha^i  \bar{s}_{j \dot{\beta} } p^{\dot{\alpha} \beta} s^j_\beta       \right)\frac{R_1^1 + R_2^2}{2}}    \nonumber \\
e^{(\bar{C}_{r}(2 q \cdot k, \bar{q} ) )^j_i R_j^i} & = & e^{ \frac{2 q^\alpha_1 k_{\alpha \dot{\alpha}}   \bar{q}^{2 \dot{\alpha} }   }{1 + 2q^\alpha_2 k_{\alpha \dot{\alpha}}   \bar{q}^{2 \dot{\alpha} }   } R^1_2   } e^{ \log\left( \frac{(1 + 2q^\alpha_2 k_{\alpha \dot{\alpha}}   \bar{q}^{2 \dot{\alpha} })^2   }{ 1 +   2q^\alpha_i k_{\alpha \dot{\alpha}}  \bar{q}^{i \dot{\alpha}} +   2 (q^\alpha_i k_{\alpha \dot{\alpha}}  \bar{q}^{i \dot{\alpha}})^2 +  2q^\alpha_i k_{\alpha \dot{\alpha}}  \bar{q}^{i \dot{\beta}} q^\beta_j k_{\beta \dot{\beta}}  \bar{q}^{j \dot{\alpha} }    }  \right) \frac{R^2_2 - R^1_1}{2}   }   e^{ \frac{2 q^\alpha_2 k_{\alpha \dot{\alpha}}   \bar{q}^{1 \dot{\alpha} }   }{1 + 2q^\alpha_2 k_{\alpha \dot{\alpha}}   \bar{q}^{2 \dot{\alpha} }   } R^2_1   }      \times    \nonumber \\
& & e^{- \log\left( 1 -   2q^\alpha_i k_{\alpha \dot{\alpha}}  \bar{q}^{i \dot{\alpha}} +   2 (q^\alpha_i k_{\alpha \dot{\alpha}}  \bar{q}^{i \dot{\alpha}})^2 -  2q^\alpha_i k_{\alpha \dot{\alpha}}  \bar{q}^{i \dot{\beta}} q^\beta_j k_{\beta \dot{\beta}}  \bar{q}^{j \dot{\alpha} }       \right)\frac{R_1^1 + R_2^2}{2}}
\end{eqnarray}
which means
\begin{eqnarray}
e^{(\bar{C}_{r}(2 q \cdot k, \bar{q} ) )^j_i R_j^i} e^{r^2_1 R_2^1} |\psi_0\rangle & = & \left( 1 -   2q^\alpha_i k_{\alpha \dot{\alpha}}  \bar{q}^{i \dot{\alpha}} +   2 (q^\alpha_i k_{\alpha \dot{\alpha}}  \bar{q}^{i \dot{\alpha}})^2 -  2q^\alpha_i k_{\alpha \dot{\alpha}}  \bar{q}^{i \dot{\beta}} q^\beta_j k_{\beta \dot{\beta}}  \bar{q}^{j \dot{\alpha} }       \right)^{-r} \times \nonumber \\
& & \left( \frac{(1 + 2 r^2_1 q^\alpha_2 k_{\alpha \dot{\alpha}}   \bar{q}^{1 \dot{\alpha} } + 2q^\alpha_2 k_{\alpha \dot{\alpha}}   \bar{q}^{2 \dot{\alpha} })^2   }{ 1   + 2q^\alpha_i k_{\alpha \dot{\alpha}}  \bar{q}^{i \dot{\alpha}} +   2 (q^\alpha_i k_{\alpha \dot{\alpha}}  \bar{q}^{i \dot{\alpha}})^2 +  2q^\alpha_i k_{\alpha \dot{\alpha}}  \bar{q}^{i \dot{\beta}} q^\beta_j k_{\beta \dot{\beta}}  \bar{q}^{j \dot{\alpha} }    }  \right)^R e^{(r')_1^2 R_2^1} |\psi_0\rangle   \nonumber \\
(r')_1^2 & \equiv & \frac{r^2_{1}(1 + 2 q_1^\alpha k_{\alpha \dot{\alpha}} \bar{q}^{1 \dot{\alpha}}   )  +  2 q_1^\alpha k_{\alpha \dot{\alpha}} \bar{q}^{2 \dot{\alpha}}   }{1 + 2 q_2^\alpha k_{\alpha \dot{\alpha}} \bar{q}^{2 \dot{\alpha}}  +  2 r^2_1 q_2^\alpha k_{\alpha \dot{\alpha}} \bar{q}^{1 \dot{\alpha}} }    \label{rpDef}
\end{eqnarray}
We are now in a position to follow on from (\ref{Master2}) which gives
\begin{eqnarray}
& & \langle   \psi' | e^{\bar{r}_i^j R^i_j} e^{ \bar{C}_{d}( \bar{s}, 2 p\cdot s ) D  } e^{ \bar{C}_{r}( \bar{s}, 2 p\cdot s )^j_i R_j^i  } e^{\bar{s}_{i \dot{\alpha}}'' \bar{S}^{i \dot{\alpha}}  }   e^{(C_r(s'' , q''  ))_i^j R^i_j }  e^{ C_d(s'', q''  ) D}  e^{   (\bar{q}'')^{i \dot{\alpha} } \bar{Q}_{i \dot{\alpha} }  } e^{ \bar{C}_{r}( 2 q \cdot k, \bar{q} )^j_i R_j^i  } e^{ \bar{C}_{d}( 2 q \cdot k, \bar{q} )  D  } e^{r_i^j R^i_j}| \psi' \rangle      \nonumber \\
&\rightarrow & \langle   \psi' | e^{r^1_2 R_1^2} e^{ \bar{C}_{d}( \bar{s}, 2 p\cdot s ) D  } e^{ \bar{C}_{r}( \bar{s}, 2 p\cdot s )^j_i R_j^i  } e^{\bar{s}_{i \dot{\alpha}}'' \bar{S}^{i \dot{\alpha}}  }   e^{(C_r(s'' , q''  ))_i^j R^i_j }  e^{ C_d(s'', q''  )D }  e^{   (\bar{q}'')^{i \dot{\alpha} } \bar{Q}_{i \dot{\alpha} }  }  e^{ \bar{C}_{r}( 2 q \cdot k, \bar{q} )^j_i R_j^i  } e^{ \bar{C}_{d}( 2 q \cdot k, \bar{q} )  D  } e^{r^2_1 R_2^1}| \psi'  \rangle   \nonumber \\
& = & \left( 1 -   2q^\alpha_i k_{\alpha \dot{\alpha}}  \bar{q}^{i \dot{\alpha}} +   2 (q^\alpha_i k_{\alpha \dot{\alpha}}  \bar{q}^{i \dot{\alpha}})^2 -  2q^\alpha_i k_{\alpha \dot{\alpha}}  \bar{q}^{i \dot{\beta}} q^\beta_j k_{\beta \dot{\beta}}  \bar{q}^{j \dot{\alpha} }       \right)^{-\frac{\Delta}{2}-r} \times \nonumber \\
& & \left( 1 -   2 \bar{s}_{i \dot{\alpha}}  p^{\dot{\alpha} \alpha } s_\alpha^i   +   2 ( \bar{s}_{i \dot{\alpha}} p^{\dot{\alpha} \alpha } s_\alpha^i  )^2 -  2 \bar{s}_{i \dot{\beta}} p^{ \dot{\alpha} \alpha } s_\alpha^i    \bar{s}_{j \dot{\alpha} } p^{ \dot{\beta} \beta} s_\beta^j        \right)^{-\frac{\Delta}{2}-r}   \nonumber \\
& &   \left( \frac{(1 + 2 r^2_1 q^\alpha_2 k_{\alpha \dot{\alpha}}   \bar{q}^{1 \dot{\alpha} } + 2q^\alpha_2 k_{\alpha \dot{\alpha}}   \bar{q}^{2 \dot{\alpha} })^2   }{ 1   + 2q^\alpha_i k_{\alpha \dot{\alpha}}  \bar{q}^{i \dot{\alpha}} +   2 (q^\alpha_i k_{\alpha \dot{\alpha}}  \bar{q}^{i \dot{\alpha}})^2 +  2q^\alpha_i k_{\alpha \dot{\alpha}}  \bar{q}^{i \dot{\beta}} q^\beta_j k_{\beta \dot{\beta}}  \bar{q}^{j \dot{\alpha} }    }  \right)^R \times \nonumber \\
& & \left( \frac{(1 + 2 r^1_2 \bar{s}_{1 \dot{\alpha} }  p^{ \dot{\alpha} \alpha} s_\alpha^2    + 2 \bar{s}_{2 \dot{\alpha} } p^{ \dot{\alpha} \alpha} s_\alpha^2   )^2   }{ 1   + 2 \bar{s}_{i \dot{\alpha}} p^{ \dot{\alpha} \alpha} s_\alpha^i  +   2 ( \bar{s}_{i \dot{\alpha}} p^{ \dot{\alpha} \alpha} s_\alpha^i  )^2 +  2 \bar{s}_{i \dot{\beta}} p^{ \dot{\alpha} \alpha} s_\alpha^i   \bar{s}_{j \dot{\alpha} } p^{ \dot{\beta} \beta} s_\beta^j     }  \right)^R   \times \nonumber \\
& & \langle   \psi' | e^{(r')^1_2 R_1^2} \left( e^{\bar{s}_{i \dot{\alpha}}'' \bar{S}^{i \dot{\alpha}}  }   e^{(C_r(s'' , q''  ))_i^j R^i_j }  e^{ C_d(s'', q''  )D}  e^{   (\bar{q}'')^{i \dot{\alpha} } \bar{Q}_{i \dot{\alpha} }  }    \right)  e^{(r')^2_1 R_2^1}| \psi'  \rangle    \nonumber 
\end{eqnarray}
We evaluate the remaining expectation value using the swap rule (\ref{SbQbSwap}) as well as an $SU(2)$ decomposition for the R-charge generator.  The final expression we obtain is
\begin{eqnarray}
& & \langle \psi_{0} | e^{r^1_2 R_1^2} e^{ \bar{l}_{\dot{2}}^{\dot{1}} \bar{L}^{\dot{2}}_{\ \dot{1}} }  e^{ l^{2}_1 L_{2}^{\ 1}} e^{ \bar{s}_{i \dot{\alpha}} \bar{S}^{i \dot{\alpha}}  } e^{ s_\alpha^i S^\alpha_i }   e^{ k_{\dot{\alpha} \alpha} K^{\alpha \dot{\alpha}} }   e^{ p^{\alpha \dot{\alpha}} P_{\dot{\alpha} \alpha} } e^{q^\alpha_i Q^i_\alpha} e^{\bar{q}^{i \dot{\alpha}} \bar{Q}_{i \dot{\alpha}}  } e^{ l^{1}_2 L_{1}^{\ 2}} e^{ \bar{l}_{\dot{1}}^{\dot{2}} \bar{L}^{\dot{1}}_{\ \dot{2}} } e^{r_1^2 R^1_2}  |\psi_{0}\rangle  \nonumber \\
&=&  \left( (1 - 4 k_{1 \dot{\beta} } p^{\dot{\beta} 1}) (1 - 4 k_{2 \dot{\beta} } p^{\dot{\beta} 2})  - 16 k_{1 \dot{\beta} } p^{\dot{\beta} 2} k_{2 \dot{\gamma} } p^{\dot{\gamma} 1} \right)^{-\Delta}  \times \nonumber \\
& &  \left(   \frac{(1 - 4 k_{1 \dot{\beta} } p^{\dot{\beta} 1} - s_1^i q_i^1) +  l^2_1 (4k_{2 \dot{\beta} } p^{\dot{\beta} 1}  +  s_2^i q_i^1)  +   l^1_2 ( 4 k_{1 \dot{\beta} } p^{\dot{\beta} 2}  +  s_1^i q_i^2  ) +    l^2_1  l^1_2 (1 - 4 k_{2 \dot{\beta} } p^{\dot{\beta} 2} - s_2^i q_i^2)       }{\sqrt{   (1 - 4 k_{1 \dot{\beta} } p^{\dot{\beta} 1}   -  s_1^i q_i^1) (1 - 4 k_{2 \dot{\beta} } p^{\dot{\beta} 2} - s_2^i q_i^2)  - (4 k_{1 \dot{\beta} } p^{\dot{\beta} 2} +   s_1^i q_i^2  )(4 k_{2 \dot{\beta} } p^{\dot{\beta} 1} +   s_2^i q_i^1   )   }  }      \right)^{2 h }  \times \nonumber   \\
& &   \left(   \frac{(1 - 4  \tilde{p}^{\dot{2} \beta} \tilde{k}_{\beta \dot{2}} - \bar{s}_{i \dot{2}}\bar{q}^{i \dot{2}}   ) + \bar{l}^{\dot{1}}_{
\dot{2}} ( 4 \tilde{p}^{\dot{2} \beta} \tilde{k}_{\beta \dot{1}} +  \bar{s}_{i \dot{1}}\bar{q}^{i \dot{2}}  )   +    \bar{l}^{\dot{2}}_{
\dot{1}} ( 4 \tilde{p}^{\dot{1} \beta} \tilde{k}_{\beta \dot{2}} +   \bar{s}_{i \dot{2}}\bar{q}^{i \dot{1}}   ) +    \bar{l}^{\dot{1}}_{
\dot{2}}  \bar{l}^{\dot{2}}_{
\dot{1}} (1 - 4 \tilde{p}^{\dot{1} \beta} \tilde{k}_{\beta \dot{1}} - \bar{s}_{i \dot{1}}\bar{q}^{i \dot{1}})       }{   \sqrt{(1 - 4  \tilde{p}^{\dot{1} \beta} \tilde{k}_{\beta \dot{1}} - \bar{s}_{i \dot{1}}\bar{q}^{i \dot{1}}) (1 - 4  \tilde{p}^{\dot{2} \beta} \tilde{k}_{\beta \dot{2}} - \bar{s}_{i \dot{2}}\bar{q}^{i \dot{2}})  - ( 4 \tilde{p}^{\dot{2} \beta} \tilde{k}_{\beta \dot{1}}  + \bar{s}_{i \dot{1}}\bar{q}^{i \dot{2}}) (4  \tilde{p}^{\dot{1} \beta} \tilde{k}_{\beta \dot{2}}  +   \bar{s}_{i \dot{2}}\bar{q}^{i \dot{1}}   )    }   }      \right)^{2 \bar{h}} \nonumber \\
& & \left( 1 -   2q^\alpha_i k_{\alpha \dot{\alpha}}  \bar{q}^{i \dot{\alpha}} +   2 (q^\alpha_i k_{\alpha \dot{\alpha}}  \bar{q}^{i \dot{\alpha}})^2 -  2q^\alpha_i k_{\alpha \dot{\alpha}}  \bar{q}^{i \dot{\beta}} q^\beta_j k_{\beta \dot{\beta}}  \bar{q}^{j \dot{\alpha} }       \right)^{-\frac{\Delta}{2}-r} \times \nonumber \\
& & \left( 1 -   2 \bar{s}_{i \dot{\alpha}} p^{\dot{\alpha} \alpha } s_\alpha^i   +   2 ( \bar{s}_{i \dot{\alpha}} p^{\dot{\alpha} \alpha } s_\alpha^i )^2 -  2 \bar{s}_{i \dot{\beta}} p^{ \dot{\alpha} \alpha } s_\alpha^i \bar{s}_{j \dot{\alpha} }  p^{ \dot{\beta} \beta} s_\beta^j        \right)^{-\frac{\Delta}{2}-r}   \nonumber \\
& &   \left( \frac{(1 + 2 r^2_1 q^\alpha_2 k_{\alpha \dot{\alpha}}   \bar{q}^{1 \dot{\alpha} } + 2q^\alpha_2 k_{\alpha \dot{\alpha}}   \bar{q}^{2 \dot{\alpha} })^2   }{ 1   + 2q^\alpha_i k_{\alpha \dot{\alpha}}  \bar{q}^{i \dot{\alpha}} +   2 (q^\alpha_i k_{\alpha \dot{\alpha}}  \bar{q}^{i \dot{\alpha}})^2 +  2q^\alpha_i k_{\alpha \dot{\alpha}}  \bar{q}^{i \dot{\beta}} q^\beta_j k_{\beta \dot{\beta}}  \bar{q}^{j \dot{\alpha} }    }  \right)^R \times \nonumber \\
& & \left( \frac{(1 + 2 r^1_2  \bar{s}_{1 \dot{\alpha} } p^{ \dot{\alpha} \alpha} s_\alpha^2   + 2 \bar{s}_{2 \dot{\alpha} } p^{ \dot{\alpha} \alpha} s_\alpha^2   )^2   }{ 1   + 2 \bar{s}_{i \dot{\alpha}} p^{ \dot{\alpha} \alpha} s_\alpha^i  +   2 (  \bar{s}_{i \dot{\alpha}} p^{ \dot{\alpha} \alpha} s_\alpha^i  )^2 +  2 \bar{s}_{i \dot{\beta}} p^{ \dot{\alpha} \alpha} s_\alpha^i  \bar{s}_{j \dot{\alpha} } p^{ \dot{\beta} \beta} s_\beta^j     }  \right)^R   \times \nonumber \\ 
& & \left( (1 + (s'')^1_\alpha (q'')_1^\alpha)(1 + (s'')^2_\alpha (q'')_2^\alpha) - (s'')^1_\alpha (q'')_2^\alpha (s'')^2_\beta (q'')_1^\beta     \right)^{r - \frac{\Delta}{2}}   \nonumber \\
& & \left( (1 + (\bar{s}'')_{1 \dot{\alpha}} (\bar{q}'')^{1 \dot{\alpha}})(1 + (\bar{s}'')_{2 \dot{\alpha}} (\bar{q}'')^{2 \dot{\alpha}}) - (\bar{s}'')_{1 \dot{\alpha}} (\bar{q}'')^{2 \dot{\alpha} } (\bar{s}'')_{2 \dot{\beta}} (\bar{q}'')^{1 \dot{\beta}}     \right)^{r + \frac{\Delta}{2}}   \nonumber \\
& & \left[ \  (1 + (\bar{s}'')_{2 \dot{\alpha}} (\bar{q}'')^{2 \dot{\alpha}} + (s'')^1_\alpha (q'')_1^\alpha )   +   (1 + (\bar{s}'')_{1 \dot{\alpha}} (\bar{q}'')^{1 \dot{\alpha}} + (s'')^2_\alpha (q'')_2^\alpha )(r')_1^2 (r')_2^1  \right. \nonumber \\
& & \left. + ( (\bar{s}'')_{1 \dot{\alpha}} (\bar{q}'')^{2 \dot{\alpha}} - (s'')^2_\alpha (q'')_1^\alpha ) (r')_2^1  +  ((\bar{s}'')_{2 \dot{\alpha}} (\bar{q}'')^{1 \dot{\alpha}} - (q'')^1_\alpha (q'')_2^\alpha ) (r')_1^2  \  \right]^{2R}  \nonumber \\
& & \hspace{-2.0cm} \left(  (1 + (\bar{s}'')_{1 \dot{\alpha}} (\bar{q}'')^{1 \dot{\alpha}} + (s'')^2_\alpha (q'')_2^\alpha )(1 + \bar{s}_{2 \dot{\alpha}} \bar{q}^{2 \dot{\alpha}} + (s'')^1_\alpha (q'')_1^\alpha ) - (\bar{s}_{1 \dot{\alpha}} \bar{q}^{2 \dot{\alpha}} - (s'')^2_\alpha (q'')_1^\alpha )(\bar{s}_{2 \dot{\alpha}} \bar{q}^{1 \dot{\alpha}} - (s'')^1_\alpha (q'')_2^\alpha )     \right)^{-R}   \nonumber
\end{eqnarray}  
where we have defined
\begin{eqnarray}
(r')_1^2 & = & \frac{r^2_{1}(1 + 2 q_1^\alpha k_{\alpha \dot{\alpha}} \bar{q}^{1 \dot{\alpha}}   )  +  2 q_1^\alpha k_{\alpha \dot{\alpha}} \bar{q}^{2 \dot{\alpha}}   }{1 + 2 q_2^\alpha k_{\alpha \dot{\alpha}} \bar{q}^{2 \dot{\alpha}}  +  2 r^2_1 q_2^\alpha k_{\alpha \dot{\alpha}} \bar{q}^{1 \dot{\alpha}} }    \nonumber \\
(r')_2^1 & = & \frac{r^1_{2}(1 + 2 \bar{s}_{1 \dot{\alpha}} p^{ \dot{\alpha} \alpha  } s^{1 }_{\alpha}   )  +  2 \bar{s}_{2 \dot{\alpha}} p^{ \dot{\alpha} \alpha  } s^{1 }_{\alpha}   }{1 + 2 \bar{s}_{2 \dot{\alpha}} p^{ \dot{\alpha} \alpha  } s^{2 }_{\alpha}  +  2 r^1_2 \bar{s}_{1 \dot{\alpha}} p^{ \dot{\alpha} \alpha  } s^{2 }_{\alpha} }    \nonumber \\
(s'')^i_\alpha (q'')^\alpha_j & = & \frac{(1 - 4 k_{1 \dot{\alpha}} p^{\dot{\alpha} 1} ) s^i_1 q^1_j + 4 k_{1 \dot{\alpha}} p^{\dot{\alpha} 2} s^i_2 q^1_j + 4 k_{2 \dot{\alpha}} p^{\dot{\alpha} 1} s^i_1 q^2_j   +  (1 - 4 k_{2 \dot{\alpha}} p^{\dot{\alpha} 2} ) s^i_2 q^2_j}{ (1 - 4 k_{1 \dot{\alpha}} p^{\dot{\alpha} 1} )(1 - 4 k_{2 \dot{\alpha}} p^{\dot{\alpha} 2} ) - 16 k_{2 \dot{\alpha}} p^{\dot{\alpha} 1} k_{1 \dot{\alpha}} p^{\dot{\alpha} 2} }    \nonumber \\
(\bar{s}'')_{i \dot{\alpha}} (\bar{q}'')^{j \dot{\alpha}} & = & \frac{(1 - 4 p^{\dot{1} \alpha} k_{\dot{\alpha} \dot{1}}) \bar{C}_{\bar{s}}(\bar{s}, 2 p\cdot s)_{i \dot{1}} \bar{C}_{\bar{q}}( 2 q\cdot k, \bar{q})^{j \dot{1}}   +  (1 - 4 p^{\dot{2} \alpha} k_{\dot{\alpha} \dot{2}}) \bar{C}_{\bar{s}}(\bar{s}, 2 p\cdot s)_{i \dot{2}} \bar{C}_{\bar{q}}( 2 q\cdot k, \bar{q})^{j \dot{2}}    }{(1 - 4 p^{\dot{1} \alpha} k_{\dot{\alpha} \dot{1}})(1 - 4 p^{\dot{2} \alpha} k_{\dot{\alpha} \dot{2}}) - p^{\dot{1} \alpha} k_{\dot{\alpha} \dot{2}} p^{\dot{2} \beta} k_{\dot{\beta} \dot{1}} }   \nonumber \\
& & + \frac{4 p^{\dot{1} \alpha} k_{\dot{\alpha} \dot{2}} \bar{C}_{\bar{s}}(\bar{s}, 2 p\cdot s)_{i \dot{1}} \bar{C}_{\bar{q}}( 2 q\cdot k, \bar{q})^{j \dot{2}}   +  4 p^{\dot{2} \alpha} k_{\dot{\alpha} \dot{1}} \bar{C}_{\bar{s}}(\bar{s}, 2 p\cdot s)_{i \dot{2}} \bar{C}_{\bar{q}}( 2 q\cdot k, \bar{q})^{j \dot{1}}    }{(1 - 4 p^{\dot{1} \alpha} k_{\dot{\alpha} \dot{1}})(1 - 4 p^{\dot{2} \alpha} k_{\dot{\alpha} \dot{2}}) - p^{\dot{1} \alpha} k_{\dot{\alpha} \dot{2}} p^{\dot{2} \beta} k_{\dot{\beta} \dot{1}} }    \nonumber \\
\bar{C}_{\bar{s}}(\bar{s}, 2 p\cdot s)_{i \dot{\alpha}} \bar{S}^{i \dot{\alpha}}  & = & \frac{ (1 -   2 \bar{s}_{j \dot{1}} p^{\dot{1} \beta } s_\beta^j) \bar{s}_{i \dot{2}} \bar{S}^{i \dot{2}}  +  (1 -   2 \bar{s}_{j \dot{2}} p^{\dot{2} \beta } s_\beta^j) \bar{s}_{i \dot{1}} \bar{S}^{i \dot{1}} + 2\bar{s}_{j \dot{1}} p^{\dot{2} \beta } s_\beta^j \bar{s}_{k \dot{2} }\bar{S}^{k \dot{1}} + 2\bar{s}_{j \dot{2}} p^{\dot{1} \beta } s_\beta^j \bar{s}_{k \dot{1} }\bar{S}^{k \dot{2}}   }{(1 -   2 \bar{s}_{j \dot{1}} p^{\dot{1} \beta } s_\beta^j)(1 -   2 \bar{s}_{j \dot{2}} p^{\dot{2} \beta } s_\beta^j)  - 4 \bar{s}_{j \dot{1}} p^{\dot{2} \alpha } s_\alpha^j \bar{s}_{k \dot{2}} p^{\dot{1} \beta } s_\beta^k  }    \nonumber \\
\bar{C}_{\bar{q}}( 2 q\cdot k, \bar{q})^{i \dot{\alpha}} \bar{Q}_{i \dot{\alpha}}  & = & \frac{ (1 -   2 q^{\alpha}_i k_{\alpha \dot{1} } \bar{q}^{i \dot{1}}) q^{j \dot{2}}\bar{Q}_{j \dot{2} } + (1 -   2 q^{\alpha}_i k_{\alpha \dot{2} } \bar{q}^{i \dot{2}}) \bar{q}^{j \dot{1}}\bar{Q}_{j \dot{1}}   +    2 q^{\alpha}_i k_{\alpha \dot{1} } \bar{q}^{i \dot{2}} \bar{q}^{j \dot{1}}\bar{Q}_{j \dot{2} } +   2 q^{\alpha}_i k_{\alpha \dot{2} } \bar{q}^{i \dot{1}} q^{j \dot{2}}\bar{Q}_{j \dot{1} } }{(1 -   2 q^{\alpha}_i k_{\alpha \dot{1} } \bar{q}^{i \dot{1}})(1 -   2 q^{\alpha}_i k_{\alpha \dot{2} } \bar{q}^{i \dot{2}})  -  4 q^{\alpha}_i k_{\alpha \dot{1} } \bar{q}^{i \dot{2}} q^{\beta}_j k_{\beta \dot{2} } \bar{q}^{j \dot{1}} }     \nonumber \\
\tilde{k}_{ \alpha \dot{\alpha }} & = & k_{ \alpha \dot{\alpha}} - \frac{1}{2}  s_{\alpha} \bar{s}_{\dot{\alpha}}   \nonumber \\
\tilde{p}^{ \dot{\alpha } \alpha } & = & p^{ \dot{\alpha } \alpha } - \frac{1}{2} \bar{q}^{\dot{\alpha}} q^{\alpha} 
\end{eqnarray}
There are, due to the Grassmann variables, many equivalent ways to write the above formulas.  Though the expressions are bulky, they are explicit.  \\ \\
The computed overlap now gives the K\"{a}hler potential after taking the logarithm.  As with all previous cases, the potential is a sum of terms with coefficients given by the quantum numbers labelling the reference state.  The super-K\"{a}hler potential fully determines the resulting superspace metric.  The involved functional dependence of the terms is likely to provide involved geodesic solutions where the development of conjugate points \cite{Balasubramanian:2021mxo} are possible.  We postpone the further study of the resulting geometry to future work.  

\subsection{Higher dimensions and $\mathcal{N} > 2$} 

We have highlighted the cases of no supersymmetry, $\mathcal{N}=1$ and $\mathcal{N}=2$ supersymmetry in four dimensions.  These cases are simpler since they allow for simple $SU(2)$ decomposition formulas for both the rotation and R-charge generators.  We emphasize, however, that the formulas (\ref{SQSwap}), (\ref{SbQbSwap}) are valid for any number of supercharges in four dimensions or for any even dimension with up to $\mathcal{N}=2$ supersymmetry.   One may wish to go beyond these examples in which case one may make use of the more general BCH formula we conjecture in (\ref{newBCH}).  It is our expectation that similar manipulations to those used in this paper may be used in such studies.

\section{Discussion \label{discussion}}

The computation of circuit complexity involves the action of unitary gates on a chosen reference state in order to obtain a desired target state.  The space of accessible target states relies on both the total number of generators and the subset of generators that transform the reference state non-trivially.  As such it is important to include as many generators as possible and study reference states that transform non-trivially under these, since this gives rise to the largest possible space of target states.  \\ \\
On a technical level, the circuits described are, in fact, generalized coherent states.  The FS metric can be computed directly from the overlap of two of these coherent states and, as such, the coherent state overlap is a central object in this approach.  By construction it is the expectation value of the circuit unitary operators with respect to the reference state.  In this paper we have highlighted the Baker-Campbell-Hausdorff formulae as powerful computational tools to, in principle, compute these expectation values.  \\ \\
Compact, closed-form expressions for the BCH formulas are, however, not always known.  In this paper we have conjectured such a closed form BCH formula (\ref{newBCH}), provided the conditions (\ref{Conditions}) hold.  For its applications inside this paper one can prove its validity in these specific cases.  This allowed us to compute super-coherent state overlaps schematically for $\mathbf{\mathfrak{su}}(2,2|\mathcal{N})$ and explicitly for $\mathcal{N} = 0, 1, 2$.  As stated, these overlaps may be used to compute the FS metric.  Indeed, for the choices of reference states used in this paper, the logarithm of these overlaps are precisely the K\"{a}hler potentials sourcing the FS metric.  \\ \\
We have postponed a detailed study of the resulting geometries to future work.  The quantum numbers labelling the reference state give rise to the sum of different terms in the K\"{a}hler potential.  The metric inherits this sum structure so that the geometry is a rather involved sum of terms.  It is plausible that these can give rise to conjugate points in the circuit complexity geometry.  On a related note, it would be fascinating to study holographic nature of $\mathfrak{su}(2,2|\mathcal{N})$ circuit complexity through the AdS/CFT dictionary.   \\ \\
The structure of the superconformal algrebra (in arbitrary dimensions and for any number of supercharges) allow for profitable use of the conjectured BCH formula.  We expect that this formula will be applicable to other cases as well and may be applied successfully in future calculations of circuit complexity.  Furthermore, BCH formulas have been used in studies of Krylov- and spread complexity \cite{Caputa:2021sib,Balasubramanian:2022tpr}.  As such, this formula may also find application in these studies.  
\\
\vspace{2.0cm}

\begin{centerline} 
{\bf Acknowledgements}
\end{centerline} 
The authors would like to thank Robert de Mello Koch for suggesting this project and providing valuable input and feedback.  H.J.R.vZ is supported by the "Quantum Technologies for Sustainable Development" grant from the National Institute for Theoretical and Computational Sciences (NITHECS).  Phumudzo Rabambi is supported by the NITHECS.

\bibliography{scftrefs}

\appendix

\section{Superconformal Algebra }
\label{AppendixA}

In four dimensions one may use the bi-spinor notation for the conformal algebra \cite{Eberhardt:2020cxo, Bianchi:2019sxz}.  Note that, throughout this article, we are working with the conformal algebra in Euclidean signature.  Define
\begin{eqnarray}
(\sigma^\mu)_{\alpha \dot{\alpha}} & = & (i I, \vec{\sigma}) \nonumber \\
(\bar{\sigma}^\mu)_{\dot{\alpha} \alpha} & = & (-i I, \vec{\sigma}) \nonumber
\end{eqnarray}
where the $su(2)$ indices are raised and lowered by 
\begin{equation}
X^a = \epsilon^{ab} X_b \ \ \ , \ \ \ X_a = \epsilon_{ab} X^b
\end{equation}
Using these one now defines 
\begin{eqnarray}
P_{\alpha \dot{\alpha}} & = &  (\sigma^\mu)_{\alpha \dot{\alpha}} P_\mu \nonumber \\
K^{\dot{\alpha} \alpha} & = &  (\bar{\sigma}^\mu)^{\dot{\alpha} \alpha} K_\mu \nonumber \\
L_{\alpha}^{ \ \beta} & = & -\frac{1}{4} (\bar{\sigma}^\mu)^{\dot{\alpha} \beta} (\sigma^\nu )_{\alpha \dot{\alpha}} L_{\mu\nu} \nonumber \\
L^{\dot{\alpha}}_{ \ \dot{\beta}} & = & -\frac{1}{4} (\bar{\sigma}^\mu)^{\dot{\alpha} \alpha} (\sigma^\nu )_{\alpha \dot{\beta}} L_{\mu\nu} 
\end{eqnarray}
In this notation the conformal algebra generators are packaged as
\begin{eqnarray}
\left[ L_{\alpha}^{\ \beta}, L_{\gamma}^{\ \delta} \right] &=& \delta_\gamma^{\ \beta} L_{\alpha}^{\ \delta} - \delta_\alpha^{\ \delta} L_{\gamma}^{\ \beta} \nonumber   \\ 
\left[ L^{\dot{\alpha} }_{\ \dot{\beta}}, L^{\dot{\gamma}}_{\ \dot{\delta}}  \right] & = & -\delta^{\dot{\gamma}}_{\ \dot{\beta}} L^{\dot{\alpha}}_{\ \dot{\delta}} + \delta^{\dot{\alpha}}_{\ \dot{\delta}} L^{\dot{\gamma}}_{ \ \dot{\beta}} \nonumber \\
\left[ L_{\alpha}^{\ \beta}, P_{\gamma \dot{\gamma} } \right] & = & \delta_{\gamma}^{\ \beta} P_{\alpha \dot{\gamma}} - \frac{1}{2}\delta_{\alpha}^{\ \beta} P_{\gamma \dot{\gamma}} \nonumber \\
\left[ L^{\dot{\alpha}}_{\ \dot{\beta}}, P_{\gamma \dot{\gamma} } \right] & = & \delta^{\dot{\alpha} }_{\ \dot{\gamma}} P_{\gamma \dot{\beta}} - \frac{1}{2}\delta^{\dot{\alpha}}_{\ \dot{\beta}} P_{\gamma \dot{\gamma}} \nonumber \\
\left[ L_{\alpha}^{\ \beta}, K^{\dot{\gamma} \gamma } \right] & = & - \delta_{\alpha}^{\ \gamma} K^{\dot{\gamma} \beta} + \frac{1}{2}\delta_{\alpha}^{\ \beta} K^{\dot{\gamma} \gamma} \nonumber \\
\left[ L^{\dot{\alpha}}_{\ \dot{\beta}}, K^{\dot{\gamma} \gamma } \right] & = & -\delta^{\dot{\gamma} }_{\ \dot{\beta}} K^{\dot{\alpha} \gamma } + \frac{1}{2}\delta^{\dot{\alpha}}_{\ \dot{\beta}} K^{\dot{\gamma} \gamma }   \nonumber \\
\left[ D, P_{\alpha \dot{\alpha} }\right] & = & P_{\alpha \dot{\alpha} } \nonumber \\
\left[ D, K^{\dot{\alpha} \alpha }\right] & = & -K^{\dot{\alpha} \alpha } \nonumber \\
\left[ K^{\dot{\alpha} \alpha}, P_{\beta \dot{\beta}} \right] & = & 4 \delta_{\beta}^\alpha \delta^{\dot{\alpha}}_{\dot{\beta}}D + 4 \delta_{\beta}^\alpha L^{\dot{\alpha}}_{\ \dot{\beta}} + 4 \delta^{\dot{\alpha}}_{\ \dot{\beta}} L_{\beta}^{\ \alpha}
\end{eqnarray}
The hermitian conjugation relations are (see appendix A of \cite{Bianchi:2006ti})
\begin{eqnarray}
D^\dag & = & D \nonumber \\
P_{\alpha \dot{\alpha}}^\dag &=& K^{\dot{\alpha} \alpha} \nonumber \\
(L_{\alpha}^{\ \beta})^\dag &=& L_{\beta}^{\ \alpha} \nonumber \\
(L^{\dot{\alpha}}_{\ \dot{\beta}})^\dag &=& L^{\dot{\beta}}_{\ \dot{\alpha}} \label{hermConj1}
\end{eqnarray}
 The benefit of introducing the spinor notation is that the supercharges and conformal supercharges are convenient to introduce in this notation as $Q^i_\alpha$, $\bar{Q}_{i\dot{\alpha}}$  and $S_i^\alpha$, $\bar{S}^{i\dot{\alpha}}$.  Additionally one introduces $R$-symmetry generators $R_{j}^i$.  The label $i$ runs from $i = 1, \cdots, \mathcal{N}$ and transform under a $U(\mathcal{N})$.  The new commutation relations are
\begin{eqnarray}
\left\{ Q^i_\alpha, \bar{Q}_{j \dot{\alpha}} \right\} & = & \frac{1}{2} \delta^{i}_{j} P_{\alpha \dot{\alpha}} \nonumber \\
\left\{ \bar{S}^{i\dot{\alpha}}, S_{j}^{\ \alpha} \right\} & = & \frac{1}{2} \delta^{i}_j K^{\dot{\alpha} \alpha } \nonumber \\
\left\{ Q^{i}_{\alpha}, S_{j}^\beta \right\} & = & \delta^i_j L_{\alpha}^{\ \beta} + \frac{1}{2}\delta^{i}_j \delta_{\alpha}^\beta D - \delta_{\alpha}^\beta R^i_j \nonumber \\
\left\{ \bar{S}^{i\dot{\alpha}}, \bar{Q}_{j \dot{\beta}} \right\} & = & \delta^i_j L^{\dot{\alpha}}_{\ \dot{\beta}} + \frac{1}{2}\delta^{i}_j \delta^{\dot{\alpha}}_{\dot{\beta}} D + \delta^{\dot{\alpha}}_{\dot{\beta}} R^i_j  \nonumber \\
\left[ R^{i}_j, R^k_l \right] & = & \delta^k_j R^i_l - \delta^i_l R^k_j \nonumber  \\
\left[ R^i_j, Q^k_\alpha \right] & = & \delta_j^k Q^i_\alpha - \frac{1}{4} \delta^i_j Q^k_\alpha \nonumber \\
\left[ R^i_j, \bar{Q}_{k \dot{\alpha}} \right] & = & -\delta_k^i \bar{Q}_{j \dot{\alpha}} + \frac{1}{4} \delta^i_j \bar{Q}_{k \dot{\alpha}} \nonumber \\
\left[ R^i_j, S_{k}^{ \alpha} \right] & = & -\delta_k^i S_{j}^{\alpha} + \frac{1}{4} \delta^i_j S_{k}^{ \alpha} \nonumber \\
\left[ R^i_j, \bar{S}^{k \dot{\alpha}} \right] & = & \delta_j^k \bar{S}^{i \dot{\alpha}} - \frac{1}{4} \delta^i_j \bar{S}^{k \dot{\alpha}}   \nonumber
\end{eqnarray}
as well as
\begin{eqnarray}
\left[ D, Q^{i}_\alpha \right] & = & \frac{1}{2} Q^i_\alpha \nonumber \\
\left[ D, \bar{Q}_{i\dot{\alpha}} \right] & = & \frac{1}{2} \bar{Q}_{i\dot{\alpha} } \nonumber \\
\left[ D, S_{i}^{\alpha} \right] & = & -\frac{1}{2} S_i^\alpha \nonumber \\
\left[ D, \bar{S}^{i \dot{\alpha}} \right] & = & -\frac{1}{2} \bar{S}^{i \dot{\alpha}} \nonumber \\
\left[ L_{\alpha}^{\ \beta}, Q^i_\gamma \right] & = & \delta_{\gamma}^\beta Q^{i}_\alpha - \frac{1}{2} \delta_{\alpha}^{\beta} Q^i_\gamma \nonumber \\
\left[ L_{\alpha}^{\ \beta}, S_i^\gamma \right] & = & -\delta_{\alpha}^{\gamma}S_i^\beta + \frac{1}{2} \delta_{\alpha}^\beta S_i^\gamma    \nonumber \\
\left[ L^{\dot{\alpha}}_{\ \dot{\beta}}, \bar{Q}_{i \dot{\gamma}} \right] & = & \delta^{\dot{\alpha}}_{\dot{\gamma}} \bar{Q}_{i \dot{\beta}} - \frac{1}{2} \delta^{\dot{\alpha}}_{\ \dot{\beta}} \bar{Q}_{i \dot{\gamma}} \nonumber \\
\left[ L^{\dot{\alpha}}_{\ \dot{\beta}}, \bar{S}^{i \dot{\gamma}} \right] & = & -\delta^{\dot{\gamma}}_{\dot{\beta} } \bar{S}^{i \dot{\alpha}} + \frac{1}{2} \delta_{\dot{\beta}}^{\dot{\alpha}} \bar{S}^{i \dot{\gamma}} \nonumber \\
\left[ P_{\alpha \dot{\alpha} }, S_{i}^\beta \right] & = & -2\delta_{\alpha}^{\ \beta} \bar{Q}_{i \dot{\alpha} } \nonumber \\
\left[ P_{\alpha \dot{\alpha} }, \bar{S}^{i \dot{\beta}} \right] & = & -2 \delta_{\dot{\alpha}}^{\dot{\beta} } Q^i_{\alpha}   \nonumber \\
\left[ K^{\dot{\alpha} \alpha}, \bar{Q}_{i \dot{\beta}} \right] & = & 2 \delta_{\dot{\beta}}^{\dot{\alpha}} S^{\alpha}_i   \nonumber \\
\left[ K^{\dot{\alpha} \alpha}, Q^i_{\beta} \right] & = & 2 \delta_{\beta}^{\alpha} \bar{S}^{i \dot{\alpha} }    \nonumber 
\end{eqnarray}
The additional hermiticity conditions are \cite{Eberhardt:2020cxo}
\begin{eqnarray}
(Q^i_\alpha)^\dag &=& S_{i}^\alpha \nonumber \\
(\bar{Q}_{i \dot{\alpha}})^\dag & = & \bar{S}^{i \dot{\alpha}}  \nonumber \\
(R_{i}^j)^\dag & = & R^i_j  \label{hermConj2}
\end{eqnarray}

\section{Overlap Simplifications}

In this appendix we proceed to manipulate (\ref{Ouroverlap}) by means of BCH formulas.  Remarkably, we are able to obtain the full dependence on the spins $h,\bar{h}$ without specifying the value for $\mathcal{N}$.  To make the sequence of manipulations easier to follow, we color in red the exponentials to be swapped in the following line.  We start with the swap rule (\ref{KPSwap}) and obtain
\begin{eqnarray}
& & \langle \psi_{0} | e^{\bar{r}^i_j R_i^j} e^{ \bar{l}_{\dot{2}}^{\dot{1}} \bar{L}^{\dot{2}}_{\ \dot{1}} }  e^{ l^{2}_1 L_{2}^{\ 1}} e^{ \bar{s}_{i \dot{\alpha}} \bar{S}^{i \dot{\alpha}}  } e^{ s_\alpha^i S^\alpha_i }  \textcolor{red}{e^{ k_{\alpha \dot{\alpha}} K^{\dot{\alpha} \alpha } }   e^{ p^{\dot{\alpha} \alpha } P_{\alpha \dot{\alpha} } } } e^{q^\alpha_i Q^i_\alpha} e^{\bar{q}^{i \dot{\alpha}} \bar{Q}_{i \dot{\alpha}}  } e^{ l^{1}_2 L_{1}^{\ 2}} e^{ \bar{l}_{\dot{1}}^{\dot{2}} \bar{L}^{\dot{1}}_{\ \dot{2}} } e^{r^j_i R^i_j}  |\psi_{0}\rangle    \nonumber \\
& = & \langle \psi_0 | e^{\bar{r}^j_i R^i_j}  e^{ \bar{l}_{\dot{2}}^{\dot{1}} \bar{L}^{\dot{2}}_{\ \dot{1}} }  e^{ l^{2}_1 L_{2}^{\ 1}} \textcolor{red}{e^{ \bar{s}_{i \dot{\alpha}} \bar{S}^{i \dot{\alpha}}  } e^{ s_\alpha^i S^\alpha_i }  e^{ p'^{\dot{\alpha} \alpha } P_{\alpha \dot{\alpha} } }   } e^{ d D} e^{\bar{\lambda}_{\dot{1}}^{\dot{2}} \bar{L}^{\dot{1} }_{\ \dot{2}} }   e^{\bar{\lambda}_0 \bar{L}^{\dot{1}}_{\ \dot{1} }  } e^{\bar{\lambda}_{\dot{2}}^{\dot{1}} \bar{L}^{\dot{2} }_{\ \dot{1}} } \times \nonumber \\
& & e^{\lambda_2^1 L_{1}^{\ 2}} e^{\lambda_0 L_{2}^{\ 2}} e^{\lambda_1^2 L_{2}^{\ 1}} \textcolor{red}{e^{ k'_{\alpha \dot{\alpha}} K^{\dot{\alpha} \alpha} }  e^{q^\alpha_i Q^i_\alpha} e^{\bar{q}^{i \dot{\alpha}} \bar{Q}_{i \dot{\alpha}}  }  } e^{ l^{1}_2 L_{1}^{\ 2}} e^{ \bar{l}_{\dot{1}}^{\dot{2}} \bar{L}^{\dot{1}}_{\ \dot{2}} } e^{r^j_i R^i_j} |\psi_0 \rangle     \nonumber   \\
& = & \langle \psi_0| e^{\bar{r}^j_i R^i_j}  e^{ \bar{l}_{\dot{2}}^{\dot{1}} \bar{L}^{\dot{2}}_{\ \dot{1}} }  e^{ l^{2}_1 L_{2}^{\ 1}}e^{ p'^{\dot{\alpha} \alpha } P_{\alpha \dot{\alpha} } } e^{2p'^{\dot{\alpha} \alpha} s_{i \dot{\alpha} } Q^i_\alpha }  e^{ \bar{s}_{i \dot{\alpha}} \bar{S}^{i \dot{\alpha}} }  e^{2 p'^{\dot{\alpha} \alpha} s^i_{\alpha} \bar{Q}_{i \dot{\alpha}} }   \textcolor{red}{e^{ s_\alpha^i S^\alpha_i }   }  e^{ d D} e^{\bar{\lambda}_{\dot{1}}^{\dot{2}} \bar{L}^{\dot{1} }_{\ \dot{2}} }   e^{\bar{\lambda}_0 \bar{L}^{\dot{1}}_{\ \dot{1} }  } e^{\bar{\lambda}_{\dot{2}}^{\dot{1}} \bar{L}^{\dot{2} }_{\ \dot{1}} } \times \nonumber \\
& & \textcolor{red}{e^{\lambda_2^1 L_{1}^{\ 2}}   } e^{\lambda_0 L_{2}^{\ 2}} \textcolor{red}{   e^{\lambda_1^2 L_{2}^{\ 1}}   }  \textcolor{red}{e^{q^\alpha_i Q^i_\alpha} } e^{2 k'_{\alpha \dot{\alpha}} q^\alpha_i \bar{S}^{i \dot{\alpha}}   }   e^{\bar{q}^{i \dot{\alpha}} \bar{Q}_{i \dot{\alpha}}  } e^{ 2 k'_{\alpha \dot{\alpha}} \bar{q}^{i \dot{\alpha}} S_i^\alpha      }    e^{ k'_{\alpha \dot{\alpha}} K^{\dot{\alpha} \alpha} }   e^{ l^{1}_2 L_{1}^{\ 2}} e^{ \bar{l}_{\dot{1}}^{\dot{2}} \bar{L}^{\dot{1}}_{\ \dot{2}} } e^{\bar{r}^j_i R^i_j} | \psi_0 \rangle    \nonumber \\
&=& \textcolor{red}{\langle  \psi_0 | } e^{\bar{r}^j_i R^i_j} e^{ \bar{l}_{\dot{2}}^{\dot{1}} \bar{L}^{\dot{2}}_{\ \dot{1}} }  e^{ l^{2}_1 L_{2}^{\ 1}}  \textcolor{red}{ e^{ p'^{\dot{\alpha} \alpha } P_{\alpha \dot{\alpha} } }   e^{2p'^{\dot{\alpha} \alpha} s_{i \dot{\alpha} } Q^i_\alpha }  } e^{\lambda_2^1 L_{1}^{\ 2}} e^{ \bar{s}_{i \dot{\alpha}} \bar{S}^{i \dot{\alpha}} }   e^{2 p'^{\dot{\alpha} \alpha} s^i_{\alpha} \bar{Q}_{i \dot{\alpha}} } e^{\bar{\lambda}_{\dot{1}}^{\dot{2}} \bar{L}^{\dot{1} }_{\ \dot{2}} }  e^{ s_\alpha^i S^\alpha_i + [s_\alpha^i S^\alpha_i, \lambda_2^1 L_{1}^{\ 2}] }     e^{ d D} \times \nonumber \\
& &  e^{\lambda_0 L_{2}^{\ 2}}  e^{\bar{\lambda}_0 \bar{L}^{\dot{1}}_{\ \dot{1} }  } \    e^{q^\alpha_i Q^i_\alpha + [\lambda_1^2 L_{2}^{\ 1}, q^\alpha_i Q^i_\alpha ]}  e^{\bar{\lambda}_{\dot{2}}^{\dot{1}} \bar{L}^{\dot{2} }_{\ \dot{1}} }   e^{2 k'_{\alpha \dot{\alpha}} q^\alpha_i \bar{S}^{i \dot{\alpha}}   }   e^{\bar{q}^{i \dot{\alpha}} \bar{Q}_{i \dot{\alpha}}  } e^{\lambda_1^2 L_{2}^{\ 1}} \textcolor{red}{ e^{ 2 k'_{\alpha \dot{\alpha}} \bar{q}^{i \dot{\alpha}} S_i^\alpha      }     e^{ k'_{\alpha \dot{\alpha}} K^{\dot{\alpha} \alpha} }  }  e^{ l^{1}_2 L_{1}^{\ 2}} e^{ \bar{l}_{\dot{1}}^{\dot{2}} \bar{L}^{\dot{1}}_{\ \dot{2}} } e^{ r^j_i R^i_j} \textcolor{red}{|\psi_0 \rangle }  \nonumber \\
&=& \langle \psi_0| e^{\bar{r}^j_i R^i_j} e^{ \bar{l}_{\dot{2}}^{\dot{1}} \bar{L}^{\dot{2}}_{\ \dot{1}} }  e^{ l^{2}_1 L_{2}^{\ 1}}  e^{\lambda_2^1 L_{1}^{\ 2}} \textcolor{red}{e^{ \bar{s}_{i \dot{\alpha}} \bar{S}^{i \dot{\alpha}} }   e^{2 p'^{\dot{\alpha} \alpha} s^i_{\alpha} \bar{Q}_{i \dot{\alpha}} } }  e^{\bar{\lambda}_{\dot{1}}^{\dot{2}} \bar{L}^{\dot{1} }_{\ \dot{2}} }  e^{ s_\alpha^i S^\alpha_i + [s_\alpha^i S^\alpha_i, \lambda_2^1 L_{1}^{\ 2}] }     e^{ d D} \times \nonumber \\
& &  e^{\lambda_0 L_{2}^{\ 2}}  e^{\bar{\lambda}_0 \bar{L}^{\dot{1}}_{\ \dot{1} }  } \    e^{q^\alpha_i Q^i_\alpha + [\lambda_1^2 L_{2}^{\ 1}, q^\alpha_i Q^i_\alpha ]}  e^{\bar{\lambda}_{\dot{2}}^{\dot{1}} \bar{L}^{\dot{2} }_{\ \dot{1}} } \textcolor{red}{  e^{2 k'_{\alpha \dot{\alpha}} q^\alpha_i \bar{S}^{i \dot{\alpha}}   }   e^{\bar{q}^{i \dot{\alpha}} \bar{Q}_{i \dot{\alpha}}  }    } e^{\lambda_1^2 L_{2}^{\ 1}}    e^{ l^{1}_2 L_{1}^{\ 2}} e^{ \bar{l}_{\dot{1}}^{\dot{2}} \bar{L}^{\dot{1}}_{\ \dot{2}} } e^{ r^j_i R^i_j} |\psi_0 \rangle   
\end{eqnarray}
We now make use of the swap rule (\ref{SbQbSwap}).  This yields
\begin{eqnarray}
&=& \langle \psi_0| e^{ \bar{r}^j_i R^i_j} e^{ \bar{l}_{\dot{2}}^{\dot{1}} \bar{L}^{\dot{2}}_{\ \dot{1}} }  e^{ l^{2}_1 L_{2}^{\ 1}}  e^{\lambda_2^1 L_{1}^{\ 2}} e^{ \bar{C}_{\bar{q}}( \bar{s}, 2 p\cdot s )^{i \dot{\alpha}} \bar{Q}_{i \dot{\alpha}}   } e^{ \bar{C}_{d}( \bar{s}, 2 p\cdot s ) D  } e^{ \bar{C}_{\bar{l}}( \bar{s}, 2 p\cdot s )^{\dot{\beta}}_{\dot{\alpha}} \bar{L}_{\ \dot{\beta}}^{\dot{\alpha}}   } e^{ \bar{C}_{r}( \bar{s}, 2 p\cdot s )^j_i R_j^i  } \textcolor{red}{ e^{ \bar{C}_{\bar{s}}( \bar{s}, 2 p\cdot s )_{i \dot{\alpha}} \bar{S}^{i \dot{\alpha}}   } e^{\bar{\lambda}_{\dot{1}}^{\dot{2}} \bar{L}^{\dot{1} }_{\ \dot{2}} }  } \times \nonumber \\
& & e^{ s_\alpha^i S^\alpha_i + [s_\alpha^i S^\alpha_i, \lambda_2^1 L_{1}^{\ 2}] }     e^{ d D + \lambda_0 L_{2}^{\ 2} + \bar{\lambda}_0 \bar{L}^{\dot{1}}_{\ \dot{1} }  } \    e^{q^\alpha_i Q^i_\alpha + [\lambda_1^2 L_{2}^{\ 1}, q^\alpha_i Q^i_\alpha ]}  \times    \nonumber \\
& & \textcolor{red}{   e^{\bar{\lambda}_{\dot{2}}^{\dot{1}} \bar{L}^{\dot{2} }_{\ \dot{1}} }  e^{ \bar{C}_{\bar{q}}( 2 q \cdot k, \bar{q} )^{i \dot{\alpha}} \bar{Q}_{i \dot{\alpha}}   }   } e^{ \bar{C}_{d}( 2 q \cdot k, \bar{q} ) D  } e^{ \bar{C}_{\bar{l}}( 2 q \cdot k, \bar{q} )^{\dot{\beta}}_{\dot{\alpha}} \bar{L}_{\ \dot{\beta}}^{\dot{\alpha}}   } e^{ \bar{C}_{r}( 2 q \cdot k, \bar{q} )^j_i R_j^i  } e^{ \bar{C}_{\bar{s}}( 2 q \cdot k, \bar{q} )_{i \dot{\alpha}} \bar{S}^{i \dot{\alpha}}   }  e^{\lambda_1^2 L_{2}^{\ 1}}    e^{ l^{1}_2 L_{1}^{\ 2}} e^{ \bar{l}_{\dot{1}}^{\dot{2}} \bar{L}^{\dot{1}}_{\ \dot{2}} } e^{ r^j_i R^i_j} | \psi_0 \rangle   \nonumber \\
&=& \langle \psi_0| e^{ \bar{r}^j_i R^i_j}  e^{ \bar{l}_{\dot{2}}^{\dot{1}} \bar{L}^{\dot{2}}_{\ \dot{1}} }  e^{ l^{2}_1 L_{2}^{\ 1}}  e^{\lambda_2^1 L_{1}^{\ 2}} e^{ \bar{C}_{d}( \bar{s}, 2 p\cdot s ) D  } e^{ \bar{C}_{\bar{l}}( \bar{s}, 2 p\cdot s )^{\dot{\beta}}_{\dot{\alpha}} \bar{L}_{\ \dot{\beta}}^{\dot{\alpha}}   } e^{ \bar{C}_{r}( \bar{s}, 2 p\cdot s )^j_i R_j^i  }  e^{\bar{\lambda}_{\dot{1}}^{\dot{2}} \bar{L}^{\dot{1} }_{\ \dot{2}} } \textcolor{red}{    e^{ \bar{C}_{\bar{s}}( \bar{s}, 2 p\cdot s )_{i \dot{\alpha}} \bar{S}^{i \dot{\alpha}}   +   [ \bar{C}_{\bar{s}}( \bar{s}, 2 p\cdot s )_{i \dot{\alpha}} \bar{S}^{i \dot{\alpha}} ,    \bar{\lambda}_{\dot{1}}^{\dot{2}} \bar{L}^{\dot{1} }_{\ \dot{2}}     ]   }   } \times \nonumber \\
& & \textcolor{red}{e^{ s_\alpha^i S^\alpha_i + [s_\alpha^i S^\alpha_i, \lambda_2^1 L_{1}^{\ 2}] }     e^{ d D + \lambda_0 L_{2}^{\ 2} + \bar{\lambda}_0 \bar{L}^{\dot{1}}_{\ \dot{1} }  } \    e^{q^\alpha_i Q^i_\alpha + [\lambda_1^2 L_{2}^{\ 1}, q^\alpha_i Q^i_\alpha ]}  } \times    \nonumber \\
& &   \textcolor{red}{   e^{ \left( \bar{C}_{\bar{q}}( 2 q \cdot k, \bar{q} )^{i \dot{\alpha}} \bar{Q}_{i \dot{\alpha}}   +     [ \bar{\lambda}_{\dot{2}}^{\dot{1}} \bar{L}^{\dot{2} }_{\ \dot{1}}   ,   \bar{C}_{\bar{q}}( 2 q \cdot k, \bar{q} )^{i \dot{\alpha}} \bar{Q}_{i \dot{\alpha}}  ]    \right)  } } e^{\bar{\lambda}_{\dot{2}}^{\dot{1}} \bar{L}^{\dot{2} }_{\ \dot{1}} } e^{ \bar{C}_{d}( 2 q \cdot k, \bar{q} ) D  } e^{ \bar{C}_{\bar{l}}( 2 q \cdot k, \bar{q} )^{\dot{\beta}}_{\dot{\alpha}} \bar{L}_{\ \dot{\beta}}^{\dot{\alpha}}   } e^{ \bar{C}_{r}( 2 q \cdot k, \bar{q} )^j_i R_j^i  }  e^{\lambda_1^2 L_{2}^{\ 1}}    e^{ l^{1}_2 L_{1}^{\ 2}} e^{ \bar{l}_{\dot{1}}^{\dot{2}} \bar{L}^{\dot{1}}_{\ \dot{2}} } e^{ r^j_i R^i_j}| \psi_0 \rangle  \nonumber
\end{eqnarray}
The highlighted supercharge swaps can be performed using (\ref{SQSwap}) and are of the form
\begin{eqnarray}
& & e^{\bar{s}_{i \dot{\alpha}} \bar{S}^{i \dot{\alpha}}  } e^{s^i_\alpha S_i^\alpha} e^{d D + \lambda_0 L_{2}^{\ 2} + \bar{\lambda}_0 \bar{L}^{\dot{1}}_{\ \dot{1} }} e^{q_i^\alpha Q^i_\alpha} e^{\bar{q}^{i \dot{\alpha} } \bar{Q}_{i \dot{\alpha} }   }     \nonumber \\
&=& e^{\bar{s}_{i \dot{\alpha}} \bar{S}^{i \dot{\alpha}}  } e^{s^i_\alpha S_i^\alpha}  e^{ e^{\lambda_0 L_{2}^{\ 2}}    \left(  e^{\frac{d}{2}} q_i^\alpha Q^i_\alpha   \right)  e^{- \lambda_0 L_{2}^{\ 2}}   } e^{  e^{ \bar{\lambda}_0 \bar{L}^{\dot{1}}_{\ \dot{1} }} \left(   e^{\frac{d}{2}} \bar{q}^{i \dot{\alpha} } \bar{Q}_{i \dot{\alpha} } \right) e^{- \bar{\lambda}_0 \bar{L}^{\dot{1}}_{\ \dot{1} }} } e^{d D + \lambda_0 L_{2}^{\ 2} + \bar{\lambda}_0 \bar{L}^{\dot{1}}_{\ \dot{1} }}  \nonumber \\
&\equiv& e^{\bar{s}_{i \dot{\alpha}} \bar{S}^{i \dot{\alpha}}  } e^{s^i_\alpha S_i^\alpha}  e^{   \tilde{q}_i^\alpha Q^i_\alpha    } e^{  \bar{\tilde{q}}^{i \dot{\alpha} } \bar{Q}_{i \dot{\alpha} }  } e^{d D + \lambda_0 L_{2}^{\ 2} + \bar{\lambda}_0 \bar{L}^{\dot{1}}_{\ \dot{1} }}    \nonumber \\
&=& e^{\left( C_q(s, \tilde{q}) \right)_i^\alpha Q^i_\alpha}  e^{\bar{s}_{i \dot{\alpha}} \bar{S}^{i \dot{\alpha}}  } e^{C_d(s, \tilde{q}  ) D}  e^{(C_r(s , \tilde{q}  ))_i^j R^i_j }   e^{  \bar{\tilde{q}}^{i \dot{\alpha} } \bar{Q}_{i \dot{\alpha} }  } e^{(C_l(s, \tilde{q}  ))_\beta^\alpha L_\alpha^{\ \beta}} e^{(C_s(s, \tilde{q}  ))^j_\beta S_j^\beta} e^{d D + \lambda_0 L_{2}^{\ 2} + \bar{\lambda}_0 \bar{L}^{\dot{1}}_{\ \dot{1} }}   \nonumber \\
&=& e^{\left( C_q(s, \tilde{q}) \right)_i^\alpha Q^i_\alpha} \left( e^{\bar{s}_{i \dot{\alpha}} \bar{S}^{i \dot{\alpha}}  }   e^{(C_r(s , \tilde{q}  ))_i^j R^i_j }   e^{  e^{\frac{C_d(s, \tilde{q}  )}{2} } \bar{\tilde{q}}^{i \dot{\alpha} } \bar{Q}_{i \dot{\alpha} }  }    \right) e^{C_d(s, \tilde{q}  ) D} e^{(C_l(s, \tilde{q}  ))_\beta^\alpha L_\alpha^{\ \beta}} e^{(C_s(s, \tilde{q}  ))^j_\beta S_j^\beta} e^{d D + \lambda_0 L_{2}^{\ 2} + \bar{\lambda}_0 \bar{L}^{\dot{1}}_{\ \dot{1} }}     \nonumber 
\end{eqnarray}
Putting this all together we obtain
\begin{eqnarray}
& & \langle \psi_{0} | e^{ \bar{r}^j_i R^i_j} e^{ \bar{l}_{\dot{2}}^{\dot{1}} \bar{L}^{\dot{2}}_{\ \dot{1}} }  e^{ l^{2}_1 L_{2}^{\ 1}} e^{ \bar{s}_{i \dot{\alpha}} \bar{S}^{i \dot{\alpha}}  } e^{ s_\alpha^i S^\alpha_i }   e^{ k_{ \alpha \dot{\alpha} } K^{\dot{\alpha} \alpha } }   e^{ p^{ \dot{\alpha} \alpha} P_{\alpha \dot{\alpha}  } } e^{q^\alpha_i Q^i_\alpha} e^{\bar{q}^{i \dot{\alpha}} \bar{Q}_{i \dot{\alpha}}  } e^{ l^{1}_2 L_{1}^{\ 2}} e^{ \bar{l}_{\dot{1}}^{\dot{2}} \bar{L}^{\dot{1}}_{\ \dot{2}} }  e^{ r^j_i R^i_j} |\psi_{0}\rangle   \nonumber \\
&=& \langle \psi_0| e^{ \bar{r}^j_i R^i_j} e^{ \bar{l}_{\dot{2}}^{\dot{1}} \bar{L}^{\dot{2}}_{\ \dot{1}} }  e^{ l^{2}_1 L_{2}^{\ 1}}  e^{\lambda_2^1 L_{1}^{\ 2}} e^{ \bar{C}_{d}( \bar{s}, 2 p\cdot s ) D  } e^{ \bar{C}_{\bar{l}}( \bar{s}, 2 p\cdot s )^{\dot{\beta}}_{\dot{\alpha}} \bar{L}_{\ \dot{\beta}}^{\dot{\alpha}}   } e^{ \bar{C}_{r}( \bar{s}, 2 p\cdot s )^j_i R_j^i  }  e^{\bar{\lambda}_{\dot{1}}^{\dot{2}} \bar{L}^{\dot{1} }_{\ \dot{2}} } \times \nonumber \\
& & \left( e^{\bar{s}_{i \dot{\alpha}}'' \bar{S}^{i \dot{\alpha}}  }   e^{(C_r(s'' , q''  ))_i^j R^i_j }   e^{  e^{\frac{C_d(s'', q''  )}{2} } (\bar{q}'')^{i \dot{\alpha} } \bar{Q}_{i \dot{\alpha} }  }    \right) e^{C_d(s'', q''  ) D} e^{(C_l(s'', q''  ))_\beta^\alpha L_\alpha^{\ \beta}}  e^{d D + \lambda_0 L_{2}^{\ 2} + \bar{\lambda}_0 \bar{L}^{\dot{1}}_{\ \dot{1} }}   \times    \nonumber \\
& &    e^{\bar{\lambda}_{\dot{2}}^{\dot{1}} \bar{L}^{\dot{2} }_{\ \dot{1}} } e^{ \bar{C}_{d}( 2 q \cdot k, \bar{q} ) D  } e^{ \bar{C}_{\bar{l}}( 2 q \cdot k, \bar{q} )^{\dot{\beta}}_{\dot{\alpha}} \bar{L}_{\ \dot{\beta}}^{\dot{\alpha}}   } e^{ \bar{C}_{r}( 2 q \cdot k, \bar{q} )^j_i R_j^i  }  e^{\lambda_1^2 L_{2}^{\ 1}}    e^{ l^{1}_2 L_{1}^{\ 2}} e^{ \bar{l}_{\dot{1}}^{\dot{2}} \bar{L}^{\dot{1}}_{\ \dot{2}} } e^{ r^j_i R^i_j} |\psi_0 \rangle    \label{finGenSwaps}
\end{eqnarray}
where 
\begin{eqnarray}
(\bar{s}'')_{i \dot{\alpha}} \bar{S}^{i \dot{\alpha}} & = &  \bar{C}_{\bar{s}}( \bar{s}, 2 p\cdot s )_{i \dot{\alpha}} \bar{S}^{i \dot{\alpha}}   +   [ \bar{C}_{\bar{s}}( \bar{s}, 2 p\cdot s )_{i \dot{\alpha}} \bar{S}^{i \dot{\alpha}} ,    \bar{\lambda}_{\dot{1}}^{\dot{2}} \bar{L}^{\dot{1} }_{\ \dot{2}}     ]  \nonumber \\
 (s'')_\alpha^i S^\alpha_i &=& s_\alpha^i S^\alpha_i + [s_\alpha^i S^\alpha_i, \lambda_2^1 L_{1}^{\ 2}]     \nonumber \\
(q'')^\alpha_i Q_\alpha^i & = &     e^{d D + \lambda_0 L_{2}^{\ 2}}\left(    q^\alpha_i Q^i_\alpha + [\lambda_1^2 L_{2}^{\ 1}, q^\alpha_i Q^i_\alpha ]    \right)e^{-d D - \lambda_0 L_{2}^{\ 2}}     \nonumber \\
(\bar{q}'')^{i \dot{\alpha}} \bar{Q}_{i \dot{\alpha}} & = & e^{d D + \bar{\lambda}_0 \bar{L}^{\dot{1}}_{\ \dot{1}}}     \left( \bar{C}_{\bar{q}}( 2 q \cdot k, \bar{q} )^{i \dot{\alpha}} \bar{Q}_{i \dot{\alpha}}   +     [ \bar{\lambda}_{\dot{2}}^{\dot{1}} \bar{L}^{\dot{2} }_{\ \dot{1}}   ,   \bar{C}_{\bar{q}}( 2 q \cdot k, \bar{q} )^{i \dot{\alpha}} \bar{Q}_{i \dot{\alpha}}  ]    \right) e^{-d D - \bar{\lambda}_0 \bar{L}^{\dot{1}}_{\ \dot{1}}}    \nonumber
\end{eqnarray}
We note that remarkably, the spin-piece containing the undotted generators is already fully determined.  This is due to the fact all the generators remaining inside the overlap, namely $\bar{Q}_{i \dot{\alpha}}, \bar{S}^{i \dot{\alpha}}, D, R_{j}^i$ commute with the $L_{\alpha}^{\ \beta}$.  By making use of $SU(2)$ BCH formulas we can thus compute the $h$-dependent piece as
\begin{eqnarray}
& & \langle \psi_0| e^{l^2_1 L_{2}^{\ 1}} e^{\lambda^1_2 L_{1}^{\ 2}} e^{ (C_l(s'', q''))^{\alpha}_\beta L_{\alpha}^{\ \beta } } e^{\lambda_0 L_{2}^{\ 2}}e^{\lambda_1^2 L_{2}^{\ 1}}e^{l^1_2 L_{1}^{\ 2}} |\psi_0\rangle   \nonumber \\
&=& \langle h,h| e^{l^2_1 L_{2}^{\ 1}} e^{\lambda^1_2 L_{1}^{\ 2}} e^{\Lambda^1_2(s'', q'') L_{1}^{\ 2}} e^{\Lambda_0(s'', q'') L_2^{\ 2}} e^{\Lambda^2_1(s'', q'') L_{2}^{\ 1}}  e^{\lambda_0 L_{2}^{\ 2}}e^{\lambda_1^2 L_{2}^{\ 1}}e^{l^1_2 L_{1}^{\ 2}} |h,h\rangle    \nonumber \\
&=& \langle h,h| exp\left\{ 2 \log\left( e^{-\frac{\lambda_0 + \Lambda_0}{2}}\left( l^1_2 l^2_1 + e^{\Lambda_0}(1 + l^2_1(\lambda_2^1 + \Lambda_2^1))(e^{\lambda_0}(1 + l^1_2 \lambda^2_1) + l^1_2 \Lambda^2_1     )      \right)     \right)   \right\}     |h,h\rangle \nonumber \\
&=& \left(   \frac{(1 - 4 k_{1 \dot{\beta} } p^{\dot{\beta} 1} - s_1^i q_i^1) +  l^2_1 (4k_{2 \dot{\beta} } p^{\dot{\beta} 1}  +  s_2^i q_i^1)  +   l^1_2 ( 4 k_{1 \dot{\beta} } p^{\dot{\beta} 2}  +  s_1^i q_i^2  ) +    l^2_1  l^1_2 (1 - 4 k_{2 \dot{\beta} } p^{\dot{\beta} 2} - s_2^i q_i^2)       }{\sqrt{   (1 - 4 k_{1 \dot{\beta} } p^{\dot{\beta} 1}   -  s_1^i q_i^1) (1 - 4 k_{2 \dot{\beta} } p^{\dot{\beta} 2} - s_2^i q_i^2)  - (4 k_{1 \dot{\beta} } p^{\dot{\beta} 2} +   s_1^i q_i^2  )(4 k_{2 \dot{\beta} } p^{\dot{\beta} 1} +   s_2^i q_i^1   )   }  }      \right)^{2 h}                 \nonumber
\end{eqnarray}
By rearranging the swap order we are also able to obtain the expression coming from the dotted rotation generators.  To do this we first swap the supercharge and conformal supercharge exponentials 
\begin{eqnarray}
& & e^{ \bar{l}_{\dot{2}}^{\dot{1}} \bar{L}^{\dot{2}}_{\ \dot{1}} }  e^{ l^{2}_1 L_{2}^{\ 1}} \textcolor{red}{e^{ \bar{s}_{i \dot{\alpha}} \bar{S}^{i \dot{\alpha}}  } e^{ s_\alpha^i S^\alpha_i } }  e^{ k_{\alpha \dot{\alpha} } K^{\dot{\alpha} \alpha } }   e^{ p^{\dot{\alpha} \alpha } P_{ \alpha \dot{\alpha}} }  \textcolor{red}{e^{q^\alpha_i Q^i_\alpha} e^{\bar{q}^{i \dot{\alpha}} \bar{Q}_{i \dot{\alpha}}  } } e^{ l^{1}_2 L_{1}^{\ 2}} e^{ \bar{l}_{\dot{1}}^{\dot{2}} \bar{L}^{\dot{1}}_{\ \dot{2}} }    \nonumber \\
&=& e^{ \bar{l}_{\dot{2}}^{\dot{1}} \bar{L}^{\dot{2}}_{\ \dot{1}} }  e^{ l^{2}_1 L_{2}^{\ 1}} e^{ s_\alpha^i S^\alpha_i } e^{ \bar{s}_{i \dot{\alpha}} \bar{S}^{i \dot{\alpha}}  }  e^{ (k_{\alpha \dot{\alpha} } - \frac{1}{2}  s^i_{\alpha} \bar{s}_{i \dot{\alpha}}    )K^{\dot{\alpha} \alpha  } }   e^{ \left( p^{ \dot{\alpha} \alpha } - \frac{1}{2} \bar{q}^{i \dot{\alpha}} q^\alpha_i   \right) P_{\dot{\alpha} \alpha} }   e^{\bar{q}^{i \dot{\alpha}} \bar{Q}_{i \dot{\alpha}}  } e^{q^\alpha_i Q^i_\alpha} e^{ l^{1}_2 L_{1}^{\ 2}} e^{ \bar{l}_{\dot{1}}^{\dot{2}} \bar{L}^{\dot{1}}_{\ \dot{2}} } \nonumber \\
& \equiv & e^{ \bar{l}_{\dot{2}}^{\dot{1}} \bar{L}^{\dot{2}}_{\ \dot{1}} }  e^{ l^{2}_1 L_{2}^{\ 1}} e^{ s_\alpha^i S^\alpha_i } e^{ \bar{s}_{i \dot{\alpha}} \bar{S}^{i \dot{\alpha}}  }  e^{ \tilde{k}_{\alpha \dot{\alpha}} K^{ \dot{\alpha} \alpha} }   e^{  \tilde{p}^{\dot{\alpha} \alpha } P_{\alpha \dot{\alpha} } }   e^{\bar{q}^{i \dot{\alpha}} \bar{Q}_{i \dot{\alpha}}  } e^{q^\alpha_i Q^i_\alpha} e^{ l^{1}_2 L_{1}^{\ 2}} e^{ \bar{l}_{\dot{1}}^{\dot{2}} \bar{L}^{\dot{1}}_{\ \dot{2}} }. \nonumber 
\end{eqnarray}
By performing the analogous operations to above we thus end up with an expression where the dotted rotation generators can be extracted from the overlap.    Combining these results with (\ref{finGenSwaps}) we find the following expression
\begin{eqnarray}
& & \langle \psi_{0} | e^{ \bar{r}^j_i R^i_j} e^{ \bar{l}_{\dot{2}}^{\dot{1}} \bar{L}^{\dot{2}}_{\ \dot{1}} }  e^{ l^{2}_1 L_{2}^{\ 1}} e^{ \bar{s}_{i \dot{\alpha}} \bar{S}^{i \dot{\alpha}}  } e^{ s_\alpha^i S^\alpha_i }   e^{ k_{ \alpha \dot{\alpha}} K^{\dot{\alpha} \alpha } }   e^{ p^{\dot{\alpha} \alpha}  P_{\alpha \dot{\alpha}} } e^{q^\alpha_i Q^i_\alpha} e^{\bar{q}^{i \dot{\alpha}} \bar{Q}_{i \dot{\alpha}}  } e^{ l^{1}_2 L_{1}^{\ 2}} e^{ \bar{l}_{\dot{1}}^{\dot{2}} \bar{L}^{\dot{1}}_{\ \dot{2}} }  e^{ r^j_i R^i_j} |\psi_{0}\rangle   \nonumber \\
&=& \left( (1 - 4 k_{1 \dot{\beta} } p^{\dot{\beta} 1}) (1 - 4 k_{2 \dot{\beta} } p^{\dot{\beta} 2})  - 16 k_{1 \dot{\beta} } p^{\dot{\beta} 2} k_{2 \dot{\gamma} } p^{\dot{\gamma} 1} \right)^{-\Delta}  \times \nonumber \\
& &  \left(   \frac{(1 - 4 k_{1 \dot{\beta} } p^{\dot{\beta} 1} - s_1^i q_i^1) +  l^2_1 (4k_{2 \dot{\beta} } p^{\dot{\beta} 1}  +  s_2^i q_i^1)  +   l^1_2 ( 4 k_{1 \dot{\beta} } p^{\dot{\beta} 2}  +  s_1^i q_i^2  ) +    l^2_1  l^1_2 (1 - 4 k_{2 \dot{\beta} } p^{\dot{\beta} 2} - s_2^i q_i^2)       }{\sqrt{   (1 - 4 k_{1 \dot{\beta} } p^{\dot{\beta} 1}   -  s_1^i q_i^1) (1 - 4 k_{2 \dot{\beta} } p^{\dot{\beta} 2} - s_2^i q_i^2)  - (4 k_{1 \dot{\beta} } p^{\dot{\beta} 2} +   s_1^i q_i^2  )(4 k_{2 \dot{\beta} } p^{\dot{\beta} 1} +   s_2^i q_i^1   )   }  }      \right)^{2 h }  \times \nonumber   \\
& &  \left(   \frac{(1 - 4  \tilde{p}^{\dot{2} \beta} \tilde{k}_{\beta \dot{2}} - \bar{s}_{i \dot{2}}\bar{q}^{i \dot{2}}   ) + \bar{l}^{\dot{1}}_{
\dot{2}} ( 4 \tilde{p}^{\dot{2} \beta} \tilde{k}_{\beta \dot{1}} +  \bar{s}_{i \dot{1}}\bar{q}^{i \dot{2}}  )   +    \bar{l}^{\dot{2}}_{
\dot{1}} ( 4 \tilde{p}^{\dot{1} \beta} \tilde{k}_{\beta \dot{2}} +   \bar{s}_{i \dot{2}}\bar{q}^{i \dot{1}}   ) +    \bar{l}^{\dot{1}}_{
\dot{2}}  \bar{l}^{\dot{2}}_{
\dot{1}} (1 - 4 \tilde{p}^{\dot{1} \beta} \tilde{k}_{\beta \dot{1}} - \bar{s}_{i \dot{1}}\bar{q}^{i \dot{1}})       }{   \sqrt{(1 - 4  \tilde{p}^{\dot{1} \beta} \tilde{k}_{\beta \dot{1}} - \bar{s}_{i \dot{1}}\bar{q}^{i \dot{1}}) (1 - 4  \tilde{p}^{\dot{2} \beta} \tilde{k}_{\beta \dot{2}} - \bar{s}_{i \dot{2}}\bar{q}^{i \dot{2}})  - ( 4 \tilde{p}^{\dot{2} \beta} \tilde{k}_{\beta \dot{1}}  + \bar{s}_{i \dot{1}}\bar{q}^{i \dot{2}}) (4  \tilde{p}^{\dot{1} \beta} \tilde{k}_{\beta \dot{2}}  +   \bar{s}_{i \dot{2}}\bar{q}^{i \dot{1}}   )    }   }      \right)^{2 \bar{h}} \times \nonumber \\
& &   \langle   \psi' | e^{\bar{r}_i^j R^i_j} e^{ \bar{C}_{d}( \bar{s}, 2 p\cdot s ) D  } e^{ \bar{C}_{r}( \bar{s}, 2 p\cdot s )^j_i R_j^i  } e^{\bar{s}_{i \dot{\alpha}}'' \bar{S}^{i \dot{\alpha}}  }   e^{(C_r(s'' , q''  ))_i^j R^i_j } e^{C_d(s'', q''  ) }  e^{   (\bar{q}'')^{i \dot{\alpha} } \bar{Q}_{i \dot{\alpha} }  }    e^{ \bar{C}_{r}( 2 q \cdot k, \bar{q} )^j_i R_j^i  } e^{ \bar{C}_{d}( 2 q \cdot k, \bar{q} ) D  } e^{r_i^j R^i_j}| \psi'  \rangle  \nonumber \\
& & \label{Master2}
\end{eqnarray}
where the state $|\psi'\rangle$ transforms as the state $|\psi_0\rangle$ under the various generators, but has $h=\bar{h} =0$.  In terms of the labels (\ref{refStateExplicit}) it is
\begin{equation}
    |\psi'\rangle = |\Delta, 0, 0; 0, 0; \left\{ R\right\} \rangle
\end{equation}The expectation value in the first line carries the $R$-charge dependent piece of the super-coherent state overlap as well the $\Delta$-dependence that depends on the the Grassmann variables. The functions involving Grassmann variables should be expanded to the relevant order depending on the value of $\mathcal{N}$ though the way they are written above is a compact way of capturing the dependence on the Grassmann variables.

\section{A BCH formula example}
\label{RecursionAppendix}

As a simple example utilising the general formulas (\ref{newBCH}) consider $SO(2,1)$.  The algebra is
\begin{eqnarray}
\left[ K, P \right] & = & 2 D \nonumber \\
\left[ D, K\right] & = & - K \nonumber \\
\left[ D, P\right] & = & P  \nonumber 
\end{eqnarray}
By making the choice the choice $A = -i \alpha^{*} K$ and $B = i \alpha P$ we satisfy the conditions (\ref{Conditions}).  We can make an appropriate ansatz for the form of, for example, $C_i$ which gives rise to a recursive relation.  These read
\begin{eqnarray}
C_{i} & = & c_i D \nonumber \\
\Rightarrow C_{i+1} & = & 2 c_{i} \alpha \alpha^* D \nonumber
\end{eqnarray}
where we used $C_{i+1} = [[A, C_i], B]$.  This relation is solved by 
\begin{equation}
C_{i} = (2 \alpha \alpha^*)^i D    \nonumber
\end{equation}
and implies that 
\begin{eqnarray}
A_{j} & = & -i \alpha^* (2\alpha \alpha^*)^{j-1} K   \nonumber \\
B_{j} & = & i \alpha  (2\alpha \alpha^*)^{j-1} P \nonumber
\end{eqnarray}
Putting everything together we have
\begin{eqnarray}
\sum_{i=1}^{\infty} \frac{1}{2^{i-1} i} C_{i} & = & -2 \log\left( 1 - \alpha \alpha^*\right) \nonumber \\ 
\sum_{i=1}^{\infty} \frac{1}{2^{i-1}} B_{i} & = &  i\frac{\alpha}{1 - \alpha \alpha^*} P \nonumber \\
\sum_{i=1}^{\infty} \frac{1}{2^{i-1}} A_{i} & = &  -i\frac{\alpha^*}{1 - \alpha \alpha^*} K   \nonumber
\end{eqnarray}
which is consistent with known identities.  \\ \\
When deriving the identities (\ref{SQSwap}), (\ref{SbQbSwap}) and (\ref{KPSwap}) we have found it most efficient to set up the recursive relations for $A_{j}$ or $B_j$.

\end{document}